\documentclass[aps,twocolumn]{revtex4-1}
\usepackage{hyperref}
\usepackage{color,helvet,times}
\usepackage{setspace} 
\usepackage{amsmath}
\usepackage{amssymb}
\usepackage{graphicx}
\usepackage{epsfig}
\pdfoutput=1

\usepackage{diagrams}
\usepackage{chemarrow}

\newcommand{\md}{d\kern-0.035cm\char39\kern-0.08cm}
\newcommand{\mL}{L\kern-0.15cm\char39}
\newcommand{\eq}[1]{\ref{#1}}

\newcommand{\dG}{\Delta G}

\newcommand{\el}{\ell}

\begin{document}
\title{Mathematical model of alternative mechanism of 
telomere length maintenance}

\author{Richard \surname{Koll\'ar}}
\email{Corresponding author: kollar@fmph.uniba.sk}
\address{Department of Applied Mathematics and Statistics, 
Faculty of Mathematics, Physics, 
and Informatics, Comenius University, Mlynsk{\'a} dolina, 
842 48 Bratislava, Slovakia}

\author{Katar{\'\i}na \surname{Bo{\md}ov\'a}}
\address{Department of Applied Mathematics and Statistics, 
Faculty of Mathematics, Physics, 
and Informatics, Comenius University, Mlynsk{\'a} dolina, 
842 48 Bratislava, Slovakia}
\address{Institute of Science and Technology, 
Am Campus 1, 3400 Klosterneuburg, Austria}

\author{Jozef \surname{Nosek}}
\address{Department of Biochemistry, Faculty of Natural Sciences, 
Comenius University, Mlynsk{\'a} dolina, 842 15 Bratislava, Slovakia}

\author{{\mL}ubom{\'\i}r \surname{Tom\'a{\v s}ka}}
\address{Department of Genetics, Faculty of Natural Sciences, 
Comenius University, Mlynsk{\'a} dolina, 842~15 Bratislava, Slovakia}

\begin{abstract}
Biopolymer length regulation is a complex process that involves a large 
number of biological, chemical, and  physical subprocesses acting 
simultaneously across multiple spatial and temporal scales. An illustrative 
example important for genomic stability is the length regulation of 
telomeres---nucleo-protein structures at the ends of linear chromosomes 
consisting of tandemly repeated DNA sequences and a specialized set of proteins.
Maintenance of telomeres is often facilitated by the enzyme 
telomerase but, particularly in telomerase-free systems, the maintenance 
of chromosomal termini depends on alternative lengthening of telomeres 
(ALT) mechanisms mediated by recombination.
Various linear and circular DNA structures were identified to participate 
in ALT, however, dynamics of the whole process is still poorly understood. 
We propose a chemical kinetics model of ALT with kinetic rates systematically 
derived from the biophysics of DNA diffusion and looping. The reaction system is 
reduced to a coagulation-fragmentation system 
by quasi-steady state approximation.
The detailed treatment of kinetic rates yields explicit formulae for expected 
size distributions of telomeres that demonstrate 
the key role played by the J-factor, 
a quantitative measure of bending of polymers. 
The results are in agreement with experimental data 
and point out interesting phenomena: an appearance of very long telomeric 
circles if the total telomere density exceeds a critical value (excess mass) and  
a nonlinear response of the telomere size distributions to the amount 
of telomeric DNA in the system. The results can 
be of general importance for understanding dynamics of telomeres in 
telomerase-independent systems as this mode of telomere maintenance 
is similar to the situation in tumor cells lacking telomerase activity.  
Furthermore, due to its universality,  the model may also serve as a prototype 
of an interaction between linear and circular DNA structures in various settings. 
\end{abstract}
 
\pacs{}

\keywords{telomeric circles; DNA looping; polymerization; excess mass}

\maketitle

\section{Introduction}
Polymerization of monomers and cyclization of polymers are well studied 
problems with numerous applications: protein folding \cite{Lapidus}, 
chromatin fiber wrapping and stretching \cite{langschiessel}, general 
intramolecular reactions in polymers \cite{perico},  rings in micelles 
\cite{wittmer}, rings in magnetic powders or beads \cite{bennaim}, etc. 
The universal biophysical principles of dynamics of polymerization and 
cyclization play also an important role in DNA length regulation, 
particularly in length maintenance of telomeric DNA 
\cite{mceachern2000telomeres}.
Although the problem of telomere length maintenance in mammalian 
cells regulated by the enzyme telomerase attracted a significant interest 
in mathematical modeling 
\cite{AKW, levy1992telomere,deboer98, Rubelj99, arkus, olofsson2010modeling, 
Peskin} (see also the recent study \cite{daoduc}), we are not aware of any study 
of  the length maintenance in a telomerase-free environment. 
Therefore we design and analyze a mathematical model of an alternative 
telomere length maintenance (i) in a telomerase independent system and  
(ii) on a time scale much shorter than the time scale of cell division. 
The model is based on analysis of linear mitochondrial DNA (mtDNA) 
found in several yeast species that represents a natural telomerase-free system 
\cite{nosek1995linear,  tomaska2004alternatives, rycovska2004linear, 
nosek2005amplification, tomaska2000,kosa2006}, and it 
is systematically built up from biophysics of local interactions of telomeres. 
The results are also of an independent interest as the system provides 
an excellent example of a mixture 
of circular and linear polymers that interact via diffusion 
and homologous recombination.

\subsection{Telomeres}
Telomeres are specialized nucleo-protein structures at the ends of linear DNA 
molecules involved in maintaining genomic stability 
\cite{mceachern2000telomeres}. Telomeric DNA together 
with associated proteins and RNAs plays an essential role
in processes involved in DNA maintenance, such as: protection of 
chromosomal ends against degradation; masking 
the ends against inappropriate action of DNA repair machineries; 
regulation of gene expression; and pairing of homologous chromosomes 
during meiosis \cite{mceachern2000telomeres}.  Telomeric sequences of 
nuclear chromosomes in most eukaryotes consist of short 
\emph{tandem repeat units} (here referred to as \emph{t-repeats}) forming 
\emph{telomeric arrays} (\emph{t-arrays}). 
Additionally, an important role in telomere length 
maintenance is believed to be played 
by \emph{t-circles}, extrachromosomal circular molecules  that consist solely 
of t-repeats. The end-replication problem associated 
with the replication of linear DNA molecules 
\cite{olovnikov1971principle,watson1972} causes perpetual loss of t-repeats 
from t-arrays and leads eventual senescence of the cell. 

Systematic understanding of telomere length 
maintenance in telomerase-free systems 
requires a suitable model 
\cite{cesare2010alternative, conomos2013}. 
Mitochondrial telomeres in yeast provide 
such an opportunity  as they contain both t-arrays and t-circles and 
the length of its t-repeat is significantly longer compared 
to human nuclear telomeres ($n \times 6$ base pairs (bp))  
yielding a higher resolution of the experimental data. 
Experimentally measured t-circle size distributions isolated from 
yeast mitochondria show  a significant feature; 
species with long t-repeats {\it Candida~parapsilosis} ($n \times 738$ bp), 
{\it C.~metapsilosis} ($n \times 620$ bp), and  {\it Pichia~philodendri} 
($n \times 288$ bp)  seem to have exponentially decreasing distributions, 
whereas the size distribution of t-circles in {\it C.~salmanticensis} 
with relatively short t-repeats  ($n \times 104$ bp) is not monotone 
\cite{nosek1995linear,nosek2005amplification,tomaska2000}.  
The aim of our work is to explain this phenomenon.  
The biological background of the problem 
is discussed in more detail in Section~\ref{s:Bb}.

\subsection{Overview of the model and results}
Telomeres in the form of t-arrays and t-circles  can be viewed as polymers of 
a t-repeat monomer; their  concentrations are 
denoted $t_n$ ($n \ge 0$) and $c_n$ ($n >0$), respectively, and
indexed according to the number $n$ of full t-repeats they contain. 
On a short time scale the telomere size distribution dynamics is governed by  
the coagulation-fragmentation equations \cite{AizBak,BallCarr}
\begin{eqnarray}
	\frac{dc_n}{d\tau} &=& 
		\sum_{n >m>0} 
			\left[k_{m,n-m}^{CC} c_m c_{n-m} - k_{m,n-m}^{C} c_{n}\right] \nonumber \\
		& & - \sum_{m>0}  
			\left[k_{m,n}^{CC} c_m c_n - k_{m,n}^{C} c_{m+n}\right]\nonumber \\ 
		& & - \sum_{m\geq 0} 
			\left[k_{n,m}^{CT} c_n t_m -  k_{n,m}^T t_{m+n}\right] ,
		\label{cn}\\
	\frac{dt_n}{d\tau} &= & 
		\sum_{n> m\geq 0} 
			\left[k_{n-m,m}^{CT} c_{n-m} t_m -  k_{n-m,m}^T t_n\right] \nonumber \\
 		& & - \sum_{m > 0} 
			\left[k_{m,n}^{CT} c_m t_n -  k_{m,n}^T t_{m+n}\right]\nonumber \\
		& & - \!\!\!\!\!\sum_{\substack{m\geq 0 \\ n+m\geq p\geq 0}} \!\!\!\!\!\!\!\!\!
			\left[k^{TT}_{m,n,p} t_m t_n - k^{TT}_{m+n-p,p,n}t_{m+n-p}t_p\right]\!\!.
\label{odeT1}
\end{eqnarray}
One of the main points of this work is that the kinetic rates 
$k^C, k^T, k^{CC}, k^{CT}$, and $k^{TT}$ 
can be completely characterized by biophysics of the system; i.e. 
by properties of diffusion and  looping  and local interactions of DNA 
(see Section~\ref{ss:r}).  The bimolecular association of particles can 
be expressed as a product of (i) a diffusion limited rate that characterizes
 diffusive properties of the particles, (ii) a combinatorial factor that counts 
the number of possible ways a particular product can be created from the 
two particles, and (iii) a reaction-rate limited factor that accounts for 
the correction of the rate due to an energy barrier needed to cross to create
a particular product. All these individual factors can be characterized by 
the existing biophysical theory. The reader can  find details about diffusion 
limited rates, combinatorial factors, and reaction-rate limited factors 
in Sections~\ref{ss:dlr}, \ref{ss:cf}, and \ref{sec:rlf}, respectively. 
Moreover, in self-interaction of polymers an additional multiplicative factor 
dubbed \emph{J-factor} appears in the formula for kinetic rates. 
The J-factor $j(n)$ measures the rate of formation of loops of length 
$n$ t-repeats on the DNA strand and an analogous rate for circularized polymers. 
More details about J-factor can be found in Section~\ref{ss:jf}. Furthermore,  
Section~\ref{ss:d} contains explanation and evaluation 
of dissociation kinetic rates. 

 \begin{figure}[htp] %
	\centering
	\includegraphics*[width = 0.48\textwidth]{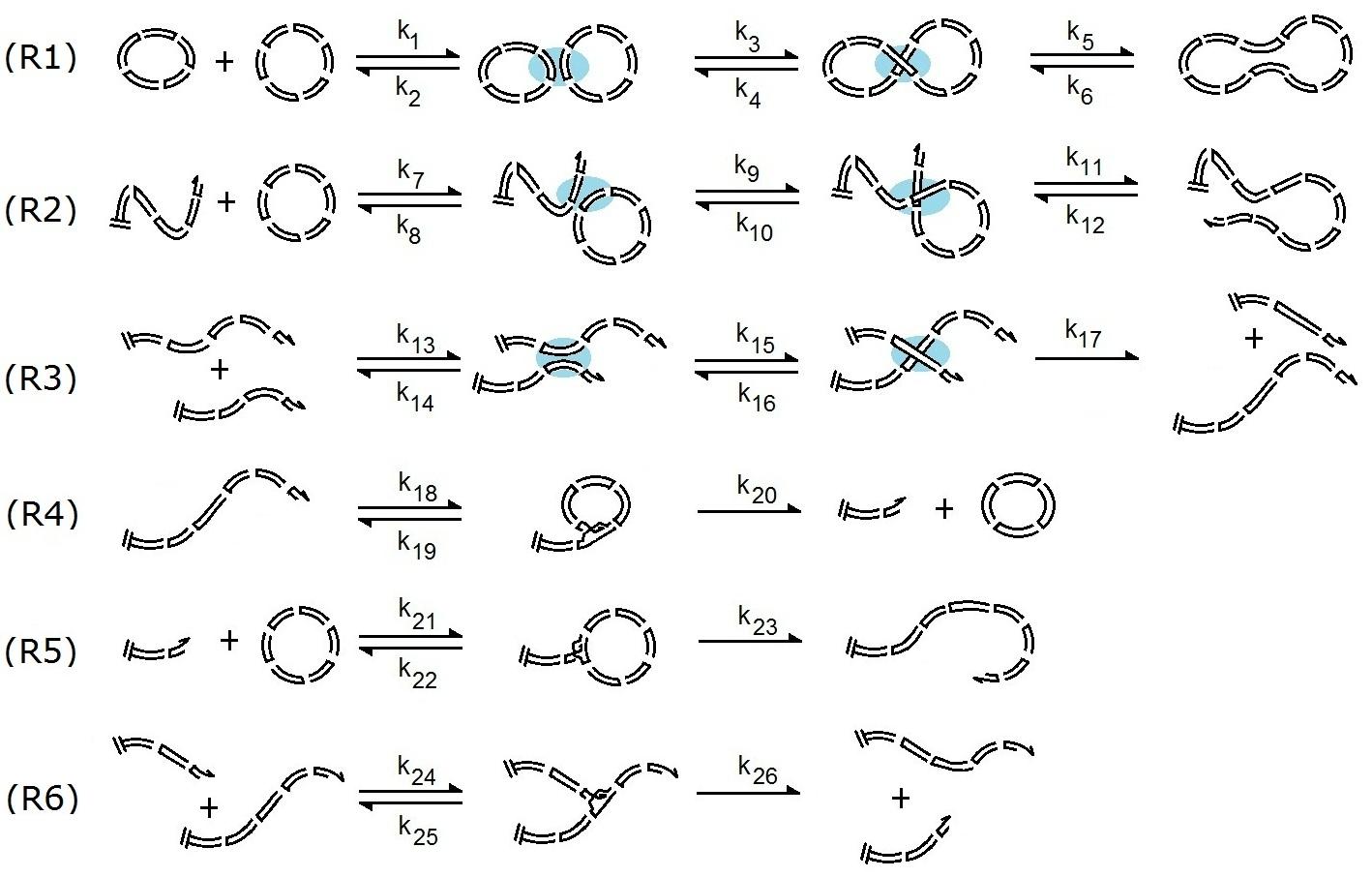}
	\caption{The full schematic list of reactions in the 
CTLY-model of ALT. The shadowed ovals represent transient recombination
complexes.  The reactions (R1)--(R3) represent recombination and 
dissociation while (R4)--(R6) represent end invasion and dissociation.}
	\label{fig_reactions}
\end{figure}

Based on the biophysical, biological, and mathematical assumptions described 
and justified in Section~\ref{ss:assum} the full list of telomere interactions 
schematically displayed on Fig.~\ref{fig_reactions} is constructed, 
see Section~\ref{ss:CTLY} for more details about this 
so-called \emph{CTLY-model}. Using the quasi-steady state approximation 
(see Section~\ref{s:CT})  the  system reduces to the \emph{CT-model} described 
by Eqs.~\ref{cn}--\ref{odeT1} where the kinetic rates are calculated from 
the biophysically derived rates for the CTLY-model (see  Appendix~\ref{sA:QSSA} for details). 
Careful bookkeeping of the composition  
of the resulting rates  in Eqs.~\ref{cn}--\ref{odeT1}  reveals that they  
satisfy detailed balance conditions and that
 the system equilibrium can be expressed by the explicit formulae
\begin{equation}
c_n = \frac{j(n)}{n} e^{-bn}\, , \ \ \ 
t_n = \frac{T}{V_0}(1-e^{-b}) e^{-bn}\, , \ \ \  b > 0,
\label{ctdist1}
\end{equation}
where $T$ is the total number of t-arrays and $V_0$ is the volume of a sample 
(see Section~\ref{s:dt}).  

\begin{widetext}
\begin{figure*}[t] %
	\centering
	\includegraphics*[width = \textwidth]{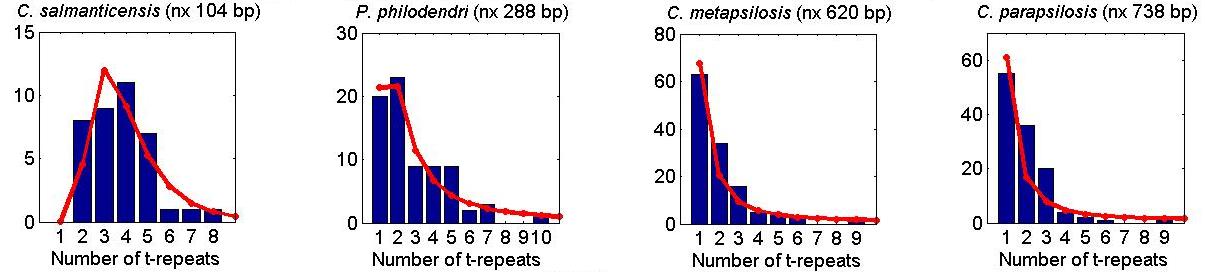}
	\caption{The comparison of the t-circle size distribution $\hat{c}_n$ 
predicted by the C-model and given by Eq.~\ref{cdist1}
with the experimental data (bar diagram). The total number of t-repeats 
and the total volume are 
$M = 141$, $V_0 = 1.43 \mu\text{m}^3$ (\emph{C.~salmanticensis}); 
$M = 217$, $V_0 = 0.32 \mu\text{m}^3$ 
(\emph{P.~philodendri}); $M = 246$, 
$V_0 = 0.5 \mu\text{m}^3$ (\emph{C.~metapsilosis}); 
and $M = 228$, $V_0 = 0.5 \mu\text{m}^3$ (\emph{C.~parapsilosis}).}
	\label{fig:fit}
\end{figure*}
\end{widetext}

The reduced \emph{C-model} that only takes into account interactions 
of the t-circles (and neglects the interactions with t-arrays) has the form 
\begin{eqnarray}
	\frac{dc_n}{d\tau} &=&
		\sum_{n >m>0} \left[k_{m,n-m}^{CC} c_m c_{n-m} - k_{m,n-m}^{C} c_{n}\right]\nonumber \\
		& & - \sum_{m>0}  \left[k_{m,n}^{CC} c_m c_n - k_{m,n}^{C} c_{m+n}\right]\, .
\label{odeC2}
\end{eqnarray}
It is introduced in Section~\ref{ss:C}. 
Its equilibrium is given by 
\begin{equation}
c_n =\frac{j(n)}{n} e^{-bn}, \qquad b \ge 0 \, .
\label{cdist1}
\end{equation}
Formulae in Eqs.~\ref{ctdist1} and \ref{cdist1} formally agree and show 
a good agreement with the experimental data (see Fig.~\ref{fig:fit}). 
However, presence of population of t-arrays 
in Eq.~\ref{ctdist1} changes the value of the parameter $b$ that may 
dramatically change the system dynamics. 
This is due to the fact that for $b \ge 0$  the reduced t-circle model 
has an equilibrium of a maximal possible mass
\begin{equation}
M_{\text{max}}  = V_0\sum_{n=1}^{\infty}  nc_n = V_0 
\sum_{n=1}^{\infty} j(n), \qquad (b = 0)\, , 
\label{Mmax}
\end{equation}
while the equilibrium in Eq.~\ref{ctdist1} admits any finite mass for $b >0$. 
Furthermore, the reduced dynamics of t-circles in Eq.~\ref{odeC2}
allows gelation effect of non-zero mass in t-circles of infinite size, while 
such a phenomenon does not appear in Eqs.~\ref{cn}--\ref{odeT1}  
(see Section~\ref{ss:em} for more explanation).

Influence of species specific  and experiment specific parameters on the telomere 
size distributions is studied in Sections~\ref{ss:ssp} and \ref{ss:esp}. 
The concluding Section~\ref{discuss} discusses open questions and possible 
utilization of  several different phenomena in applications and in 
further studies.  Section~\ref{ss:db} contains an additional interesting 
problem related to  experimental data. Finally, we point readers interested 
in analysis of the coagulation-fragmentation systems to  Appendix~\ref{sA:CT}
 where they can find a review of the literature  in the field.
We note that the kinetic rates derived for  interacting telomeres do not fit 
into any of the classes analyzed in the literature  and thus the theoretical 
studies of the systems with these rates pose an interesting open problem 
(see Appendix~\ref{sA:CT}).

\section{Biological Background}\label{s:Bb}
\subsection{Alternative Telomere Length Maintenance}
The main mechanism that prevents shortening of chromosomal termini 
is based on the reverse transcriptase activity of the enzyme \emph{telomerase}, 
the RNA-protein complex composed of the template RNA subunit  and the 
protein catalytic subunit that extends the 3' single-stranded telomeric 
overhang and thus prevents shortening of chromosomes 
\cite{greider1985identification,greider1987telomere}. However, 
telomerase-mediated synthesis of chromosomal DNA is not the only 
mechanism of telomere maintenance \cite{lundblad2002telomere}. 
Examples of telomerase-independent pathways
include (i) retrotransposition in {\it Drosophila} 
\cite{pardue2003retrotransposons},
(ii) telomeric loops (\emph{t-loops}) \cite{griffith1999mammalian}, 
(iii) chromosome circularization in mutant strains of both fission yeast 
\cite{nakamura1998two} 
and {\it Streptomyces} \cite{qin2002survival}, and in mitochondria 
of a number of yeast species \cite{nosek1995linear, rycovska2004linear}, 
(iv) heterochromatin amplification-mediated telomere maintenance (HAATI) 
mechanism \cite{jain2010haati}, 
and (v) homologous recombination. The latter was elaborated mainly 
by studies on telomere 
maintenance in yeast \cite{mceachern1996,teng1999telomere,teng2000telomerase}, 
but it was also found 
to operate in a wide variety of organisms  including some insects, 
plants \cite{biessmann1997telomere} 
and humans \cite{dunham2000telomere}. Homologous recombination 
does not only help to maintain 
telomeres, but is also involved in generation of genomic plasticity 
and instability in the absence of telomerase, 
resulting in amplification or rapid deletion of telomeric DNA and  
in formation of extrachromosomal 
telomeric fragments \cite{lundblad2002telomere, niida2000telomere}. 
Homologous recombination in telomere dynamics is considered to be  
one of the hallmarks of 
telomerase-deficient cancer cell lines maintaining their telomeres via  
\emph{Alternative Lengthening of Telomeres} 
mechanism (ALT) \cite{cesare2010alternative,tomaska2009telomeric}.

\begin{figure}[htp] %
	\centering
	\includegraphics*[width = 0.5\textwidth]{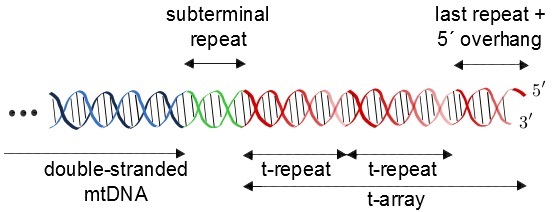}
\caption{Telomere structure of the linear mtDNA in yeast \emph{C.~parapsilosis}.}
	\label{Fig:R0}
\end{figure}

\subsection{Telomeric arrays and circles}
Linear mitochondrial genome of the yeast  {\it C.~parapsilosis}
 carrying a standard set of mitochondrial genes is represented 
by population of double-stranded DNA molecules of the length 30,923 bp 
terminating on both sides by telomeres 
consisting of a subterminal repeat (554 bp) and a t-array of t-repeats 
($n \times 738$ bp)  (Fig.~\ref{Fig:R0}). Individual molecules
in this population differ in size by the number of full-length t-repeats, 
ranging from 0 to at least 12. 
The shortest molecules (30,923 bp) terminate only with an incomplete 
tandem unit, while the termini 
of longer molecules are extended into t-arrays on both ends 
($30,923 + (n+m) \times 738$ bp, 
$m$ and $n$ are the numbers of full t-repeats in t-arrays of the 
left and right telomere, respectively).
The very ends of the molecules have 
single-stranded 5' overhangs of about 110 nucleotides 
(as opposed to 3' overhangs that are typical for nuclear DNA), 
which in a fraction of molecules invade into the double-stranded 
region (\emph{end invasion}) thus forming a duplex loop structure 
termed \emph{t-loop} \cite{tomaska2002t}.
Single-stranded overhangs are believed to be associated with 
a recombination mode of replication of telomeres 
including nuclear telomeres \cite{OganesianKarlseder}. 

Linear DNA molecules 
are accompanied by series of double-stranded circular molecules 
dubbed \emph{telomeric circles (t-circles)}. Their sizes 
correspond to integral multimers of the t-repeat with 
a length ($n \times 738$ bp). The circular molecules may 
originate from the t-loops and/or the t-arrays 
by recombination transactions followed by excision of 
a circle \cite{tomaska2000}. On the other hand, 
the t-circles were shown to replicate independently of 
the linear genomic molecules via \emph{rolling-circle mechanism} 
leading to amplification of the linear arrays of t-repeats. 
Overall, they represent a substrate for recombinational 
mode of the mitochondrial telomere 
maintenance \cite{nosek2005amplification}. Investigation 
of mutant cells lacking the t-circles revealed that 
they contain a circularized derivative of the genome. 
This supports the idea that the t-circles play a key role 
in the mechanism of telomere maintenance 
\cite{tomaska2004alternatives,kosa2006,rycovska2004linear}. 
This mechanism does not 
require telomerase activity; rather it relies on a relatively complex 
interplay among t-circles, lasso-shaped rolling-circle 
replication intermediates, t-loops and t-arrays. 

\begin{figure} 
\centering
	\includegraphics*[width = 0.48\textwidth]{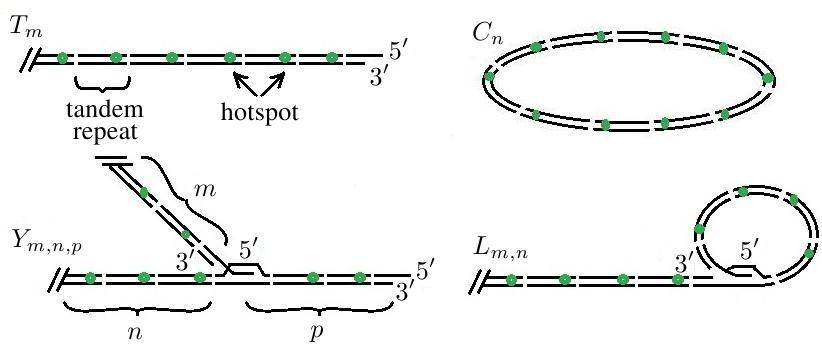}
\caption{Telomeric structures participating in replication of 
linear mtDNA containing t-repeats: 
$T_m$ a  t-array with $m$ t-repeats, and a 5' single-stranded 
overhang, $C_n$ a t-circle with $n$ t-repeats. Also displayed 
are intermediate products
$L_{m,n}$ a t-loop formed by an invasion of the 5' overhang
into the same telomeric strand of the total length $m+n$ t-repeats 
with the capping loop of the length $n$ t-repeats, and $Y_{m,n,p}$ 
a copy-choice structure (a triple junction) resembling a letter Y. 
Hotspots within each t-repeat are indicated by circles.
The skew lines at the end of $T_m$, $Y_{m,n,p}$, and $L_{m,n}$ 
represent a chromosome not displayed on the figure.}
\label{fig:1}
\end{figure}

Various simple types of telomeric structures are schematically 
illustrated on Fig.~\ref{fig:1}. Analogous telomeric structures 
(t-arrays, t-loops, t-circles) are associated with eukaryotic nuclear 
telomeres pointing to a general significance of the mitochondrial 
model \cite{tomaska2009telomeric}.

\section{Model and Data}\label{s:Models}

\subsection{Physical scales and assumptions}\label{ss:assum}
Biophysically and biologically feasible assumptions about the experimental 
environment and physical scales limit the types of structures and reactions 
involved in ALT and determine the constitution of the kinetic rates from 
biophysical components.
The telomeric structures in the model are classified according to their length 
into classes of  (see Fig.~\ref{fig:1})
\begin{itemize}
\item
 t-arrays $T_m$, $m\ge 0$, with $m$  full t-repeats and a  5' overhang, 
\item
t-circles $C_n$, $n \ge 1$, with $n$ full t-repeats. 
\end{itemize}
Two t-arrays located at different ends of a single chromosome 
are, for simplicity,  considered to be two independent particles. 

\emph{Physical scales.} 
Detailed telomere and protein structure is necessary
to unveil individual aspects of telomere length maintenance. 
However, to understand the size distributions of telomeres we focus on the 
coarse-grained spatial scale with a typical size of hundreds 
of nanometers on which we distinguish size, type, and shape of individual 
telomeres but ignore 
their nucleotide sequence. The typical time scale is set to one second, 
the time scale short for DNA degradation, cell division, and 
for a significant contribution of synthesis of new t-repeats via the 
rolling-circle mechanism.  Despite the fact that conservation of the total 
number of t-repeats valid on the fast time scale breaks down on  
a longer time scale commensurate with laboratory experiments, 
we argue that the measurements at each instance display 
``an equilibrium distribution"  for the actual number of t-repeats, 
i.e., a quasi-steady state.

\emph{Assumptions.}
The assumptions underlying the model can be divided into three classes: 
assumptions on biology and biophysics of the system, 
and mathematical assumptions. 

The assumptions on the biology are
(i) all telomeres are double-stranded with the exception 
of  a single-stranded overhang; 
(ii) more complex telomeric structures than those considered 
are dynamically unstable; 
(iii) there is only one specific reactive group, \emph{hotspot}, 
per t-repeat that is susceptible to homologous recombination; 
(iv) there is an abundance of binding proteins in the environment, 
i.e., availability of proteins does not limit local reaction rates that can 
be then viewed as crossings of energy barriers.

The assumptions (i) and (ii) were validated by 
the study \cite{tomaska2009telomeric} where no other 
telomeric structures were observed. That can be 
interpreted either as a lack of other types of structures or 
a sign of their dynamical instability. 
The assumption (iii) is not essential for our work, and in general, 
more hotspots of different structure may be active 
within one t-repeat. The structure of hotspots within one t-repeat 
only affects a single parameter of our problem ($g_1$) that may
be species dependent; it only influences the time scale of convergence 
to equilibrium telomere size distributions and
does not alter their shape. 

The assumption (iv) is typical in the Kramer's theory; 
a widely accepted theory of local chemical interactions. 
In the particular laboratory experiments that correspond to the biological 
system we study the yeast cells were held under normal physiological conditions 
and thus binding proteins are expected to be available in equilibrium 
concentrations. Also, any variation of their concentration would only 
change the values of the correction factors $g_1, g_0$ and $g_{-1}$, 
i.e., our results once again show that it has no influence 
on the shape of the resulting telomere size  distributions.

The assumptions on the system biophysics are (v) telomeres are well-mixed 
and are freely diffusing in a homogeneous three-dimensional confinement 
at a constant temperature; and (vi) telomeres do not have super-helical 
structure and do not interact with such super-coiled structures.

The assumption (v) is an assumption commonly used in chemical kinetics; 
it allows to describe the dynamics in a form of a system of ordinary 
differential equations rather than physically more realistic partial 
differential equations. In the absence of a fluid flow motion of telomeres 
in an aquatic solution is governed by diffusion and elastic forces 
within the DNA. Despite the possibility that a tangled geometry of 
mitochondria may strongly influence diffusivity and 
bending of telomeres, small dimensions of telomeres (particularly of hotspots) 
motivate us to neglect such a spatial 
inhomogeneity. The issue of possible differences in a behavior of a single 
cell vs. cell population is addressed in Section~\ref{discuss}. 

The assumption (vi) is certainly violated in the biological systems as various 
experimental studies show presence of telomeres of both relaxed and 
super-coiled geometries. Although the super-coiled molecules have different 
diffusive and reactive properties from the relaxed geometry telomeres 
considered here, they, according to the experiments, 
only account for a small fraction of telomeric structures. Here, for simplicity 
we neglect their presence and their effect on the telomere size distributions 
will be a subject of our further research. 

Obviously, in a real environment none of these biological and biophysical 
assumptions is exactly satisfied, nevertheless, 
we assume that their failure does not significantly change the qualitative 
behavior of the system.

We put forward also one mathematical assumption. We describe population 
dynamics of telomeres  by a system of ordinary differential equations that, 
in general, requires large enough populations. 
However, in our application population levels of telomeres with a large number 
of t-repeats are typically small. We believe that the good agreement of 
predictions of our model with experimental 
data provides its justification, although an extreme caution is necessary 
in interpretation of the results for populations of long telomeres.

\subsection{Reaction rates}\label{ss:r}
The repetitive structure of telomeres restricts their interactions  
to three types: homologous recombination, end invasion, 
and dissociation  (Fig.~\ref{fig_reactions}).
Homologous recombination in reactions (R1)--(R3) requires 
proximity of two identical spatially aligned hotspots on the same 
strand or on two different strands.  
Under favorable conditions (availability of recombination proteins, etc.) 
a transient complex is formed resulting either 
in an original configuration of strands or in an exchange of 
nucleotide sequences between the strands.

While homologous recombination is a reaction of two identical 
reacting groups (hotspots), 
end invasion in reactions (R4)--(R6) is an insertion of the single-stranded 
overhang located at the terminus of a DNA strand into the same or 
another strand.  Invasion into the same t-array produces a capped form 
of a telomere (t-loop) while invasion into a different t-array yields 
a copy-choice structure (triple junction) (Fig.~\ref{fig:1}). 
Again, the repetitive structure 
of the telomeres implies that each single-stranded overhang is only able 
to invade a specific location within a t-repeat 
that contains a region with identical sequence as the overhang. 

According to the Kramer's theory \cite{Jackson} even under favorable 
conditions each local interaction of telomeres 
requires crossing of an energy barrier that effectively reduces the overall 
rate by a reaction-rate limited factor.
The overall kinetic rates of individual reaction types of telomere interactions 
have the following structural decomposition:
\begin{itemize}
	\item bimolecular association of free particles
		\begin{equation*}
		r = \begin{pmatrix}  \text{diffusion} \\ \text{lim.~rate}  
\end{pmatrix} \times
				\begin{pmatrix}  \text{comb.}  \\ \text{factor}  
\end{pmatrix} \times
				\begin{pmatrix}    \text{reaction--rate}  \\ 
\text{lim.~factor} \end{pmatrix} \, ,
		\end{equation*}
	\item unimolecular self interaction (association 
of two groups of the same reactant)
		\begin{equation*}
		r = 		\begin{pmatrix}  \text{diffusion} \\ \text{lim.~rate} 
\end{pmatrix} \times
				\begin{pmatrix}  \text{comb.} \\ \text{factor} 
\end{pmatrix} \times
				\begin{pmatrix}  \text{reaction--rate} \\  
\text{lim.~factor} \end{pmatrix} \times			
				\begin{pmatrix}  \text{J-factor} \end{pmatrix}  ,
		\end{equation*}
	\item dissociation
		\begin{equation*}
		r = 		\begin{pmatrix}  \text{diffusion} \\ \text{lim.~rate} 
\end{pmatrix} \times
				\begin{pmatrix}  \text{reaction--rate} \\  
\text{lim.~factor} \end{pmatrix}  / 
				{\begin{pmatrix}  \text{reaction}\\ 
\text{volume} \end{pmatrix}}\, .
		\end{equation*}
\end{itemize}
The individual constituents are: 
(a) the diffusion limited rate that measures the rate at which two molecules 
encounter each other if driven only by a molecular diffusion; 
(b) the combinatorial factor that accounts for the number of different 
configurations of reactants in 
a reaction yielding the same product; 
(c) the reaction-rate limited factor that accommodates reaction-limited 
interactions for homologous recombination and end invasion; 
(d) the J-factor that accounts for bending of a polymer in terms of a local 
molar concentration of a reactive group at a given site; 
and (e) the reaction volume.

\subsection{Diffusion limited rate}\label{ss:dlr}
Diffusion limited kinetic rates \cite{robertson2006, Jackson}  
are determined by the size and the shape of reacting molecules.
Dependence of the rate on the polymer size  is often suppressed in 
modeling of telomere interactions \cite{olofsson2010modeling,Peskin} 
in agreement with 
the classical theory of  \citet{FloryPolymer} that suggests to neglect 
such an effect.
Hence we visualize every telomeric structure with $n$ full t-repeats as $n$ 
independent recombination hotspots each with the 
same diffusive properties of a single hotspot. Analogous view is used for the 
single-stranded overhang and corresponding matching sequences 
within each t-repeat. 
For simplicity, we do not distinguish between these two cases as we expect 
the length of the reactive group to be approximately 
the same (about 10 bp) for both recombination and end invasion.  However, 
note that the single-stranded overhang is expected to have higher mobility
(and thus different diffusive properties) than the 
double-stranded hotspot. To simplify the presentation and to 
streamline the parameter dependencies
 we include such an effect into the reaction-rate limited factor 
that condenses the various differences between 
recombination and end invasion (see Section~\ref{sec:rlf}). 

The Smoluchowski theory \cite{robertson2006} predicts the kinetic 
rate of a bimolecular reaction of two freely diffusing spherical molecules 
$A$ and $B$ to be
\begin{equation}
k_D(A,B) = 4\pi D(A,B) r(A,B)
\label{kDAB}
\end{equation}
with the joint diffusion coefficient $D(A,B) =D(A)+D(B)$ and 
the joint molecular radius $r(A,B)=r(A)+r(B)$.
The diffusion coefficient $D(A)$ of a spherical molecule $A$  
depends on its physical parameters and on the environment via 
the Stokes-Einstein formula 
\begin{equation}
D(A) = \frac{k_B T_{\text{a}}}{6\pi \hat{\nu} r(A)}\, ,
\label{DCL}
\end{equation}
where $k_B$, $T_{\text{a}}$, $\hat{\nu}$ and $r(A)$ are the 
Boltzmann constant, the absolute temperature, solvent viscosity, 
and the radius of the molecule $A$, respectively.  Equations~\ref{kDAB} 
and \ref{DCL} yield the bimolecular diffusion limited reaction 
rate of two hotspots 
\begin{equation*}
k :=  k_D(H,H) = \frac{8k_B T_{\text{a}}}{3\,\hat{\nu}}\, .
\end{equation*}
Note that the rate is independent of the dimensions of the hotspot 
indicating a universal rate for two identical molecules in agreement 
with the well-known observation of \citet{Jacobson}. 
The effective reaction volume of two interacting hotspots is given by
\begin{equation*}
V := V(H,H) = \frac{4}{3}\pi r(H)^3\, , 
\end{equation*}
where $r(H)$ is the effective radius of a hotspot $H$. 
However, in our model none of the parameters  $k$, $V$, and $r(H)$ 
has an impact on the telomere equilibrium size distribution
 as they only influence the overall time
scale of convergence of the distribution to the equilibrium, 
see Section~\ref{s:CT}.

\subsection{Combinatorial Factor}\label{ss:cf}
\begin{figure}[htp] %
	\centering
	\includegraphics*[width = 0.48\textwidth]{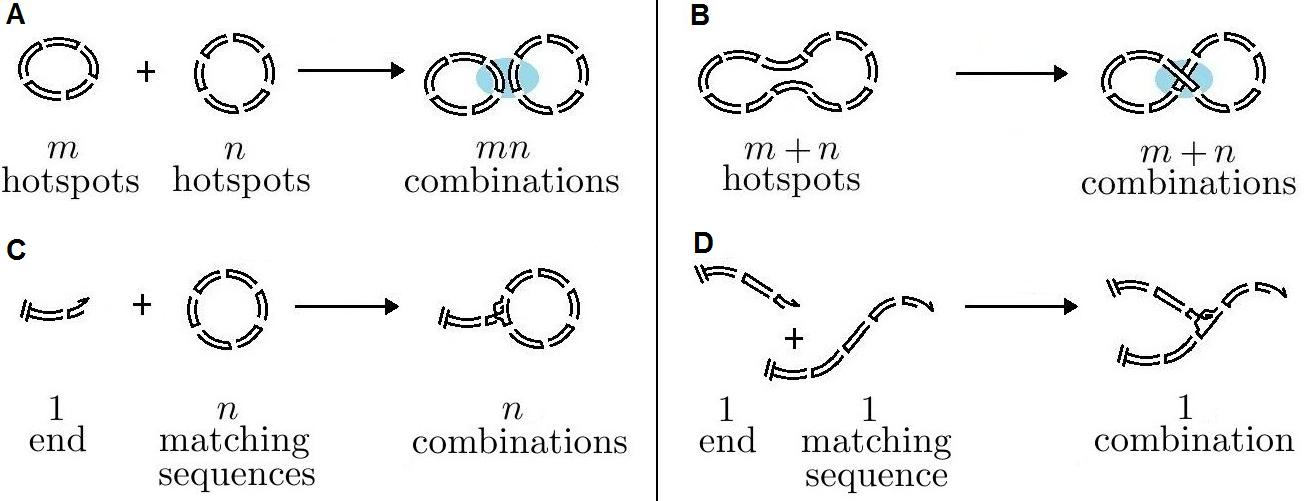}
	\caption{(A) Recombination of $C_m$ and $C_n$ yields the 
identical recombination complex for all $mn$ choices of pairs 
of interacting hotspots.
(B) Self recombination of two hotspots on $C_{m+n}$ yields 
the identical recombination complex for $m+n$ choices of pairs of 
interacting hotspots at the distance $m$ (or $n$) t-repeats along the t-circle.
(C) End invasion of $T_m$ into $C_n$ with $n$ choices. 
(D) End invasion of $T_m$ into $T_n$ with only one choice 
of the interacting site for each type of the complex.}
\label{Fig:R1}
\end{figure}

Since recombination
of any two hotspots on two different t-circles of the classes $C_m$ 
and $C_n$ (or alternatively on a t-circle $C_n$ and a t-array $T_m$) 
yields the same complex (reactions (R1) and (R2) in Fig.~\ref{fig_reactions}), 
the combinatorial factor accounting for the number of possible pairs 
of reacting hotspots  is a product $mn$ (Fig.~\ref{Fig:R1}A).
The combinatorial factor for an interaction of two recombination 
hotspots on a single t-circle $C_{m+n}$ is equal to the total number 
$m+n$ of hotspots on the t-circle for all  possible recombination 
complexes (Fig.~\ref{Fig:R1}B). 

\begin{figure}[htp] %
	\centering
	\includegraphics*[width = 0.48\textwidth]{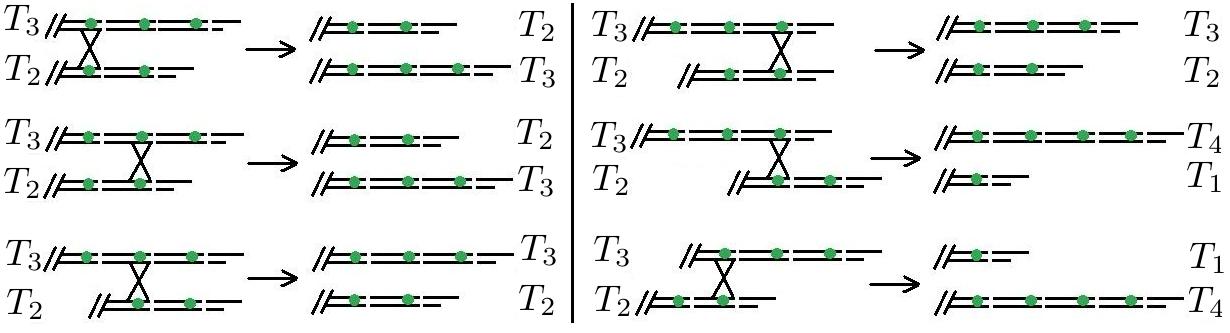}
	\caption{Recombination of t-arrays. Different outcomes 
of recombination of t-arrays $T_3$ and $T_2$ 
are schematically displayed. Two combinations lead to $T_2$ and $T_3$ 
and one combination leads to $T_1$ and $T_4$. Similarly, two 
combinations  produce $T_3$ and $T_2$ and one produces $T_4$ and $T_1$. 
In general, $\min(m,n,p,m+n-p)$ combinations 
of recombination hotspots of $T_m$ and $T_n$  result in $T_p$ and $T_{m+n-p}$.}
	\label{Fig:R3}
\end{figure}

Recombination of two hotspots on two distinct t-arrays yields
different types of recombination complexes  (Fig.~\ref{Fig:R3}). 
A simple calculation reveals that there are $\min(m,n,p,m+n-p)$ 
different pairs of hotspots on $T_m$ and $T_n$ that yield a recombination 
complex which after an exchange of ends decomposes into $T_p$ and $T_{m+n-p}$.

On the other hand, end invasion requires participation of the 
single-stranded overhang in the interaction eliminating a part of the 
combinatorial factor. Each t-repeat on the invaded telomeric strand 
has one specific matching sequence. 
The overall combinatorial factor of end invasion of a t-array into $C_n$ 
and $T_n$ is thus $n$ and 1, 
respectively (see Figs.~\ref{Fig:R1}C and \ref{Fig:R1}D).  

\subsection{Reaction-Rate Limited Factor} \label{sec:rlf}
A reaction-rate limited factor accounts for the fact that two molecules 
within the reactive distance have to overcome an additional potential barrier 
to react \cite{Jackson, phillips2009}. 
Note that this factor may also include a cumulative effect of an availability 
of connecting proteins facilitating interactions,  a presence of a capping protein 
attached to the single-stranded overhang \cite{Peskin}, and enhanced diffusion 
of single-stranded DNA.

\begin{figure} [htp]
	\centering
	\includegraphics*[width = 0.4\textwidth]{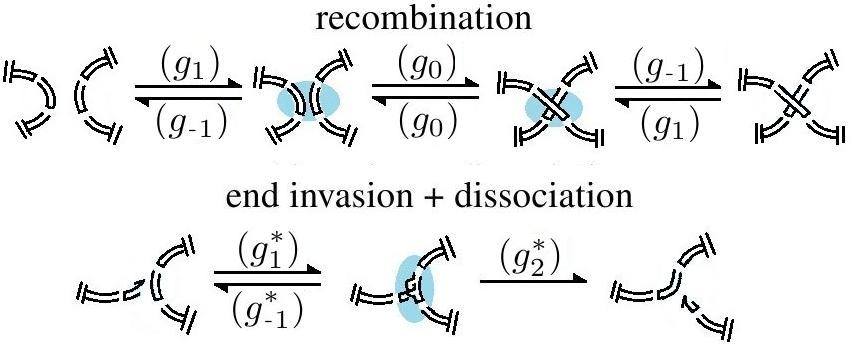}
	\caption{Schematic diagrams of local interactions: recombination, 
end invasion and dissociation 
	of  telomeric structures. The reaction-rate limited correction 
	factors are indicated.}
	\label{fig:scheme}
\end{figure}

The Noyes theory \cite{noyes1961} identifies the second order correction  
$\kappa$ that affects the diffusion-limited reaction rate and can be interpreted 
as a barrier crossing:
\begin{eqnarray}
k_N(A,B) &= &k_D(A,B) \frac{\kappa}{k_D(A,B) + \kappa} = k_D 
\left(1 + \frac{k_D}{\kappa}\right)^{-1}
\nonumber\\  
&\approx &
k_D \exp(-\triangle G/k_BT_{\text{a}}) = k_D g\, , \label{kNdef}
\end{eqnarray}
where $g:=\exp(-\triangle G/k_B T_{\text{a}})$.  Equation~\ref{kNdef} 
defines the energy 
barrier $\triangle G$ and the correction factor $g$  of a particular reaction.
The correction factors $g_1, g_0, g_{-1}$ are associated with recombination, 
and $g^{\ast}_1, g^{\ast}_2, g^{\ast}_{-1}$ 
are associated with end invasion 
(see Fig.~\ref{fig:scheme} for a schematic display). 

\subsection{J-factor -- the measure of DNA looping}\label{ss:jf}
Looping of DNA strands plays an important role in internal cell regulation 
and particularly in gene regulation.  Passive diffusive motion leads to 
a DNA strand configuration that is favorable for creation of a link between 
its two sections  (often in presence of a binding protein). The well studied 
examples are regulation of {\it Lac} repressor \cite{Towles} and {\it gal} 
operon \cite{EUreview} that block DNA transcription. In both cases, the DNA 
loop is site-specific and it does not detach from the strand. Another example 
of site-specific looping is recombination \cite{EUreview} involving {\it Cre} 
protein connecting two {\it loxP} sites that leads to an excision of a circular 
DNA fragment  that regulates gene switching. 

The problem of DNA looping has attracted a lot of attention 
in scientific community 
since the publication of the seminal paper of 
\citet{Jacobson} 
that introduced the so-called J-factor. The traditional approach is based
on the worm-like chain (WLC) polymer model of \citet{kratky1949}, although 
alternative biophysical and computational methods
are used as well. In 1980s a series of biological experiments conducted by 
\citet{Shore} and a theoretical essay by
\citet{Shimada} lied down a theoretical framework for the subject that was 
further tested in numerical
simulations. 
A qualitative difference between DNA looping {\it in vivo} and {\it in
vitro} for short strands was pointed out by \citet{Ringrose} 
who proposed an introduction of apparent persistence length for DNA polymers 
{\it in vivo} as a consequence of a presence of chromatin.
Review papers \cite{Rippe1995, EUreview, VoloIMA} provide a long 
list of literature  on looping of linear DNA. 
The theory of circular (double) constrained looping is developed in
\cite{Bloomfield1, Rippe2001, VoloMulti, HM}. 

\begin{figure}[htp] 
	\centering
	\includegraphics*[width = 0.48\textwidth]{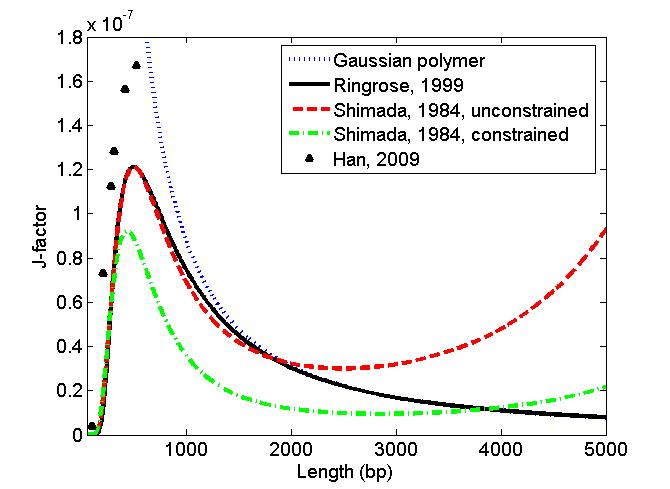}
	\caption{The values of J-factor measured in units of 
$\text{M} = 1\, \text{mol}\cdot \text{dm}^{-3}$ in various models.}
	\label{Fig:J}
\end{figure}

A disagreement in the extent of a couple of orders of magnitude was observed 
between values of $J$-factor {\it in vivo} and {\it in vitro} \cite{Han, Ringrose} 
(Fig.~\ref{Fig:J})
and between {\it in vivo} values and theoretically computed values,  
particularly for lengths
shorter than 100 bp (see the diagram \cite[Fig.~10]{Towles}).
These differences were explained in \cite{saiz2006} and  studied 
analytically and numerically in a
particular case of {\it LacI} repressing in \cite{Towles}. 
The same group conducted further experiments and compared their 
calculations with theoretical predictions 
\cite{Han}. Their results were recently supported by experimental 
study of \citet{VaHa}, see also \cite{nelson2012spare}, pointing out 
the failure of the WLC model for looping of short DNA fragments.

The linear J-factor expresses availability of a given site in a neighborhood 
of another site in terms of a local molar concentration
($\text{M} = 1 \text{mol}\cdot \text{dm}^{-3}$) 
\begin{equation}
k_{J} = j(L) k_D\, ,
\label{s8}
\end{equation}
where  $k_D$ is the kinetic rate of the appropriate bimolecular reaction 
\cite{Shore, sanchez2011}
and $L$ is the distance of the reacting hotspots along the DNA strand 
measured in bp
\footnote{The coefficient 2  in corresponding Eq.~\ref{s8} in \cite{Shore}  
accounts for two indistinguishable ends of the linear polymers while here 
the interacting hotspots are distinguishable.}.

Three types of DNA looping that may lead to an excision of a t-circle appear in 
ALT with their corresponding J-factors:
 (i) $j(L)$, linear constrained looping---recombination of 
two hotspots of a single t-array;
 (ii) $j^{\ast}(L)$, linear unconstrained looping---end invasion 
into the same t-array; 
(iii) $j(L_1,L_2)$, circular constrained  looping---recombination 
of two hotspots of a single t-circle. 
Due to the disagreement in the literature it is not completely 
clear which values of J-factor apply in the setting of telomeres in yeast mtDNA. 
The t-repeats of mtDNA of the studied species of yeast 
(with the exception of  {\it C. salmanticensis})
have lengths much longer than the persistence length of 
double-stranded DNA ($\el_p = 50$ nm)
and the lengths of t-arrays cover a wide range ($0.1$--$10$ kbp). 
Thus, for simplicity, we use the linear  J-factor values of  \citet{Ringrose} 
(Fig.~\ref{Fig:J})
\begin{equation}
j(L)  = \frac{1.25 \times 10^5}{\el_p^3} \,
\left( \frac{4\el_p}{10^4 L}\right)^{1.5}
\exp \left(\frac{-460 \,\el_p^2}{6.25 L^2}\right)\, , 
\label{jL}
\end{equation}
for both constrained and unconstrained linear looping 
($j(L)=j^{\ast}(L)$). The model in Eq.~\eq{jL} agrees for lengths 
over 2 kbp with the Gaussian polymer model 
\begin{equation}
j_{G} (L)  =  1.66 \, \left( \frac{3}{4\pi L \el_p}\right)^{1.5}\, .
\label{jG}
\end{equation}
The rate of circular looping (circular J-factor) appears in 
the backward reaction (R1) (Fig.~\ref{fig_reactions}).
The circular J-factor $j(m,n)$ measures the likelihood of looping 
of a DNA strand in situations when two particular hotspots on the 
same t-circle of a length $m+n$ t-repeats at a distance $m$ (or $n$) 
t-repeats along the t-circle form a recombination complex 
that can be upon successful recombination resolved into t-circles 
$C_m$ and $C_n$. Both lengths $m$ and $n$ contribute to the J-factor 
through the total entropy and elastic energetic loss. 
Under assumption of relaxed t-circle geometry, we use the detail 
balance condition
\cite{Rippe2001, HM}
\begin{equation}
	j(L_1, L_2) = \frac{j(L_1)j(L_2)}{j(L_1+L_2)}\, .
\label{circJ}
\end{equation}
We have considered various other alternatives for both linear and 
circular J-factor  (see Section~\ref{discuss} for more details)
but the variation either did not significantly alter the results or 
it led to a disagreement with the experimental data.

\subsection{Dissociation}\label{ss:d}
In the case of diffusion limited dissociation kinetic rates are 
determined by diffusive properties of the molecules \cite{Jackson}.  
The rate of dissociation $k_d$ of spherical 
molecules $A$ and $B$ with effective radii $r(A)$ and $r(B)$  
in the case of very weak attractive forces 
is given by 
\begin{equation*}
	k_d(A,B) =  \frac{3D(A,B)}{r(A,B)^2} = \frac{k_D(A,B)}{V(A,B)}\, .
\end{equation*}
If the interaction potential between the molecules is $U(r)$  
(the related attractive force is $d U(r) / dr$)
then the dissociation rate is given by
\begin{equation}
	k_{dU} = \frac{3D e^{U(r)/k_BT_{\text{a}}}}
{r^3 \int_{r}^{\infty}\exp(U(\rho)/k_BT_{\text{a}}) / \rho^2 \, d\rho}= 
\frac{k_D}{V} g \, ,
	\label{kDU}
\end{equation}
where arguments $(A,B)$ are suppressed. 
The correction factor $g$ can be interpreted in terms of the Kramer's 
theory as an energy barrier of height $\Delta G_d$ (see also Eq.~\ref{kNdef}):
\begin{equation*}
k_d(A,B) =  \frac{k_D(A,B)}{V(A,B)}  \exp\left(\frac{-\Delta G_d}
{k_BT_{\text{a}}}\right)\, .
\end{equation*}
Strongly diffusion limited dissociation corresponds to 
$U(r) \equiv 0$ in Eq.~\ref{kDU} and $\Delta G_d = 0$,  
analogously to strongly diffusion limited association by recombination 
($\kappa=\infty$,  Eq.~\ref{kNdef}) where $\dG = 0$.

\subsection{CTLY-model} \label{ss:CTLY}
Figure \ref{fig_reactions} schematically displays interactions 
(R1)--(R6) of telomere structures: 
t-arrays, t-circles, t-loops, triple junctions, and recombination complexes, 
in our most complex model of ALT, 
dubbed the {\it CTLY-model}.  
Although telomere structures of arbitrary length make the size 
of the system virtually infinite,
the conservation of the total number of t-repeats in the system 
effectively bounds the maximal telomere size.

Telomere interactions in the CTLY-model are homologous recombination 
of t-arrays and t-circles in reactions (R1)--(R3), 
and end invasion of a t-array into the same or another t-array or t-circle 
in (R4)--(R6) (see Fig.~\ref{fig_reactions}). 
Intermediate complexes are subsequently formed in (R1)--(R3) reflecting 
the three-stage process: 
approach of the two reactive parts, local protein mediated recombination, 
and dissociation.
Due to the assumption of limited reactivity of the intermediate complexes 
we assume that they do not self-interact or recombine, 
neither they are invaded by telomeric overhangs to form more complex 
structures (double circles, quadruple junctions, etc.). 

The rates in the CTLY-model can be directly deduced from the above discussion. 
Let $n$ and $m$ be the numbers of t-repeats in particular t-circles and t-arrays 
involved in the reactions 
in the CTLY-model displayed on Fig.~\ref{fig_reactions}.  The association 
kinetic rates are given by
\begin{eqnarray*}
k_1 & = & k_1(C_n,C_m) = k_7 = k_7(T_m,C_n) = mn k  g_1 \,, \\
k_6 & = & k_6(C_m, C_n) =  (m+n) j(n,m) k g_1\, , \\
k_{12} & = & k_{12}(T_m,C_n) = m j(n) k g_1\,,\\ 
k_{13} & = & k_{13}(T_m,T_n,T_p) = \min (m,n,p, m+n-p) k g_1\, ,\\  
k_{18} & = & k_{18}(T_m,C_n) =  j^{\ast} (n) k  g^{\ast}_1\, , \\
k_{21} & = & k_{21}(T_m,C_n) =  n k  g^{\ast}_1\, , \\
k_{24} & = & k_{24}(T_m,T_n) = k  g^{\ast}_1\, .
\end{eqnarray*}
The arguments $C_n$ and $T_m$ are specified in such a way that the reaction 
is uniquely identified although the particular molecules do not need 
to participate 
in it (as in $k_6$, the indexes $m$ and $n$ identify the lengths of the looped 
arcs created in the recombination complex). The rate $k_{13}$ requires 
the information on the product of the recombination complex $T_p$ 
(and complementary $T_{m+n-p}$). 
The dissociation rates are
\begin{gather*}
k_2 = k_5 = k_8 = k _{11} = k_{14} = k_{17} = k g_{-1}/V, \\
k_3   =   k_4  = k_9 = k_{10} = k_{15} = k_{16}= k g_0/V\, ,\\
k_{19} = k_{22} = k_{25} = k g^{\ast}_{-1}/V\, , \quad 
k_{20} = k_{23} = k_{26} = kg^{\ast}_2/V.
\end{gather*}

\subsection{Reduction to CT-model}\label{s:CT}
\emph{The quasi-steady state approximation (QSSA)} \cite{Atkins, Eigen} 
provides a reduction 
of reaction kinetics when a replacement of
some of differential equations in the system 
by their equilibria does not significantly alter the dynamics of the whole 
system on a specifically considered time scale. 
The selected components in ``equilibrium" are often called 
\emph{slaved variables} as their approximated dynamics is slaved to the 
rest of the system by an algebraic law expressing the equilibrium.  
Validity of such an approximation can be justified using the mathematical 
technique 
of singular perturbation \cite{SegelSlemrod,SM2003,TE2004} under 
an assumption of a separation of time scales,  
specifically under overall imbalance of association and dissociation. 
While the use of the QSSA can be more-or-less justified for every individual 
reaction displayed  schematically 
on Fig.~\ref{fig_reactions}, the validity needs 
to be justified for the full system of reactions simultaneously. 
Unfortunately, the existing 
theory does not cover such a case and thus it poses an interesting and
 difficult open 
mathematical problem. Nevertheless, here we apply the QSSA 
in the CTLY-model as our numerical simulations suggest 
its validity. 

The QSSA reduces the CTLY-model to the \emph{CT-model}: 
\begin{eqnarray}
	 C_m + C_n  & \autorightleftharpoons{\small $k^{CC}$}{\small $k^{C}$} &  
C_{m+n}\, , \label{eq_CC}\\
	 T_m + C_n  & \autorightleftharpoons{\small $k^{TC}$}{\small $k^{T}$} &  
T_{m+n}\, , \label{eq_CT}\\
	 T_m + T_n  & \autorightarrow{\small $k^{TT}$}{} &  T_p + T_{m+n-p}\, ,
\label{eq_TT}
\end{eqnarray}
where reaction (R1) (see Fig.~\ref{fig_reactions}) was reduced to 
Eq.~\ref{eq_CC},  reactions (R2), (R4) and (R5) 
were reduced and added up to Eq.~\ref{eq_CT}, and similarly reactions 
(R3) and (R6) were reduced and added up 
to Eq.~\ref{eq_TT}.  More details on the QSSA and on the explicit reduction to 
Eqs.~\ref{eq_CC}--\ref{eq_TT} can be found in Appendix~\ref{sA:QSSA}.

After rescaling of the time variable  $\tau = \hat{k} t g_1$ the kinetic rates 
in Eqs.~\ref{eq_CC}--\ref{eq_TT} are
\begin{eqnarray}
k^{CC}_{m,n}  & = &  k^{CC} (C_m,C_n) =   mn, \nonumber \\
k^C_{m,n} & =&  k^{C} (C_m,C_n) =  (m+n)  j(m,n) \, , \nonumber\\
k^{TC}_{m,n} & = &  k^{TC} (T_m, C_n) =  n  (m + \alpha), \label{kTCr} \\
k^T_{m,n} & = & k^{T} (T_m,C_n) = j(n)(m + \alpha)\,, \nonumber \\ 
k^{TT}_{m,n,p}  & = &  k^{TT} (T_m, T_n, T_p) =  \min(m,n,p,m+n-p) 
+ \alpha \, . \nonumber
\end{eqnarray}
The non-dimensional parameters characterizing the relative strength 
of the energy barriers of the local interactions,
and the rescaled diffusion coefficient $k$ are given by 
\begin{equation}
\beta = \frac{g_{-1}}{g_0}, \ \ 
\gamma = \frac{g_{-1}^\ast}{g_2^\ast}\, , \ \ 
\alpha = \frac{(1+\gamma) g_1^\ast}{(2+\beta)g_1}, \ \ 
\hat{k} = \frac{k}{2+\beta}.
\label{abg}
\end{equation}
The resulting (infinite) system of differential equations in 
Eqs.~\ref{cn}--\ref{odeT1} 
describes time evolution of the concentrations $c_m$ and $t_n$ of the 
populations of $C_m$ and $T_n$, $m>0, n\geq 0$, in the sample solvent.

The continuous quantities $c_m$ and $t_n$ can be interpreted as 
statistical measures (number densities) of populations of telomeres 
of type $C_m$ and $T_n$   in the sample, or (particularly in the case 
of their value smaller than one) as probabilities of an occurrence of an 
element of the given type in the sample. The actual number of particles 
of a given class is given by $\hat{t}_m = V_0 t_m$ and $\hat{c}_n = V_0 c_n$, 
where $V_0$ is the sample volume.
Important conserved quantities of the CT-model are the total number 
of t-repeats (total mass) $M$ and the total number
of t-arrays  $T$ in the system. Also,  $M= M_C + M_T$, where
$M_C$ and $M_T$ are the total numbers of t-repeats in t-circles and 
t-arrays in the sample, respectively:
\begin{equation}
M_C = \sum_{n=1}^{\infty} n\hat{c}_n\, , \ \ 
M_T = \sum_{n=1}^{\infty} n\hat{t}_n\, , \ \ 
T = \sum_{n=0}^{\infty} \hat{t}_n\, .
\label{totalmass}
\end{equation}

\subsection{Reduction to C-model}\label{ss:C}
Equations~\ref{kTCr} reveal that there is no time scale separation between 
interactions of t-circles with themselves and mutual interactions of t-circles 
and t-arrays. Nevertheless, due to the particular structure 
of Eqs~\ref{cn}--\ref{odeT1} it is  possible to study a reduction of 
Eqs.~\ref{cn}--\ref{odeT1} to account only for interactions of t-circles.  
Absence of linear telomeres ($t_n = 0$ for $n \ge 0$) reduces the dynamics
of the CT-model to the \emph{C-model} described by  Eq.~\ref{odeC2}.
The kinetic rates $k^{CC}$ and $k^{C}$ remain unchanged and they 
are given by Eqs.~\ref{kTCr}. The main advantage of this reduced C-model 
is that it fits to an existing mathematical framework of  
coagulation-fragmentation \cite{AizBak,BakBak,BallCarr,daCosta}, 
although the reaction rates do not fall directly into classes studied 
theoretically up to this date. 

\begin{figure} [h]
	\centering
	\includegraphics*[width = 0.48\textwidth]{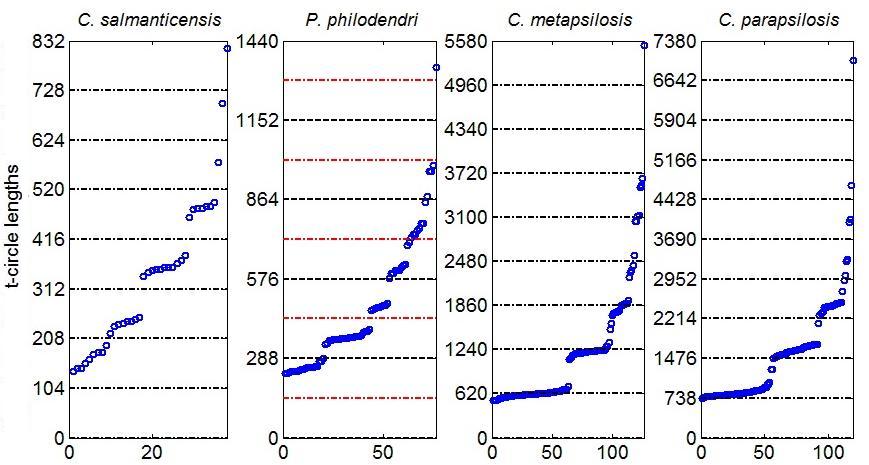}
	\caption{Apparent experimentally measured lengths of  
t-circles ordered by size (in bp).  
Multiples of sizes of t-repeats of individual yeast species  
are  indicated by horizontal lines. Additional 
lines at half of the t-repeat length are shown for {\it P.~philodendri}. 
Clustering of t-circles may be observed for all species, however, the 
gap between clusters  in case of {\it P. philodendri} is approximately 
half of the length of the t-repeat (288 bp).}
	\label{Fig:binning}
\end{figure}

\subsection{Data binning}\label{ss:db}
In reported experimental data \cite{tomaska2000} obtained by electron microscopy 
(see Appendix~\ref{sA:setup} for details) 
length of multiple t-circles was located far from its expected value 
(an integer multiple of the length of the corresponding t-repeat). 
One can attempt to categorize the t-circles according to their nearest multiple 
of the t-repeat length but 
due to the high length variability it is  difficult to assign them into a proper 
size distribution bin. 
However, the actual t-circle size distributions display an emerging feature 
illustrated on Fig.~\ref{Fig:binning}. 
They are naturally clustered to groups with well separated lengths 
corresponding to classes 
of t-circles with different integer numbers of t-repeats. Based on this natural 
clustering we have rejected the original binning method of 
 \cite{tomaska2000} and classified the t-circles according 
to the clusters recounted directly from the experimental electron microscopy 
data.  The clusters of {\it C.~metapsilosis} and {\it C.~parapsilosis} 
are relatively easy to classify with measurement precision decreasing 
as the t-circle size grows. Unfortunately,  
it is unclear how should be some of the clusters associated to the 
number of t-repeats, particularly in the case of {\it C.~salmanticensis}. 
Also, {\it P.~philodendri} with t-repeat size 288 bp has the gap between 
clusters approximately halved in contrast with other yeast species where 
the cluster gaps match the t-repeat size. 

Based on the observation of clusters of the other yeast species we 
have categorized the first cluster of the t-circles of {\it C.~salmanticensis} 
with apparent t-circle sizes between 104 bp and 208 bp as t-circles with 
two full t-repeats, i.e., we conjecture that there is no t-circle with one 
t-repeat. The main argument is that we expect that the apparent length 
of a planar microscopic image of a t-circle underestimates the actual t-circle 
length due to the two-dimensional projection and it cannot by any means 
significantly extend the expected t-repeat size. Further clusters of  t-circles 
of {\it C.~salmanticensis} are categorized accordingly.  The clusters of 
{\it P.~philodendri} were categorized as $C_1$, $C_2$, etc., 
despite the smaller cluster gap.

\section{Results}
Three models with a decreasing level of complexity were constructed
using the approach described in Section~\ref{s:Models}: the CTLY-model, 
the CT-model, and the C-model.
The most complex CTLY-model was studied only through numerical simulations 
that confirmed the validity of the QSSA. The experimental data were matched 
with the explicit equilibria of the CT-model and the C-model. 
Furthermore, we performed parametric studies of the CT-model 
to detect the system response to parameters and 
analyzed the C-model rigorously using the mathematical theory 
of coagulation-fragmentation equations.

\subsection{Distribution of t-circles}\label{s:dt}
{\it C-model.} 
The expected size distribution of the t-circles $\hat{c}_n$ with $n$ 
full t-repeats in the C-model is an equilibrium of the system of 
Eqs.~\ref{odeC2}, $n \ge 1$ 
\footnote{a further mathematical analysis reveals that the convergence 
to the equilibrium in the system is exponential}. 
Due to the structure of the kinetic rates and the fact that the circular 
J-factor satisfies the detailed balance condition 
in Eq.~\ref{circJ} the equilibrium can be explicitly calculated as 
\begin{equation}
\hat{c}_n = V_0\frac{j(n)}{n} e^{-bn}\, ,
\label{cdist}
\end{equation}
where $b \ge 0$ is a parameter that is uniquely determined by $M= M_C$. 
The quantity $M_C$ is experiment specific but it was not measured 
in experiments as the protocol does not guarantee detection of all telomeres 
in a sample. As expected the equilibrium distribution of t-circles
in this simple approximation strongly depends on the 
individual species t-repeat length (through the J-factor) with
shorter telomeres more difficult to bend and longer (pliable) telomeres 
with smaller tendency 
for circularization because of a small probability of a proper alignment of 
the specific reactive sites (entropy effects). 
Thus, according to Eq.~\ref{cdist} the J-factor appears to be  the main source 
of the difference between the character of the data for different yeast species.
A comparison of the data with the C-model predictions 
of t-circle size distributions for four distinct yeast species 
is displayed on Fig.~\ref{fig:fit}. 
The same parameters were used for matching except the individual length 
of telomeric repeat and the experiment specific total number of t-repeats 
$M$ and the sample volume~$V_0$ that were determined by minimization 
of the sum of squared error (SSE). 
All optimal values of $M$ and $V_0$ lie within an expected range.

\begin{figure}[htp]
	\centering
	\includegraphics*[width = 0.47\textwidth]{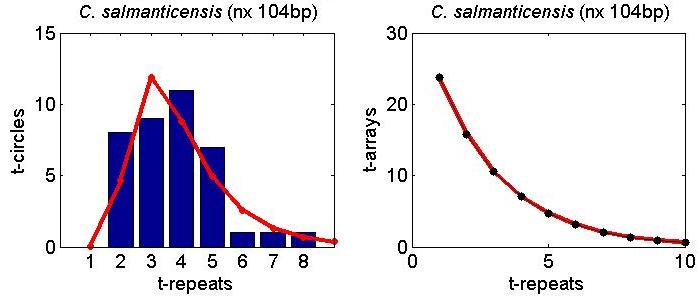}
	\caption{The distribution of t-circles and t-arrays of 
\emph{C.~salmanticensis} in the CT-model given by Eq.~\ref{ctdist}. 
While the distribution of t-circles is not monotone (the model prediction 
is compared with the bar diagram of experimental data), the size distribution 
of t-arrays is exponentially decaying. The parameters used are $T=50$, 
$V_0 = 1  \mu\text{m}^3$, and $M =292$ that reduces to $M_C = 92$.}
	\label{fig:Tarray}
\end{figure}

{\it CT-model.} 
Similarly, the equilibrium of the CT-model in Eqs.~\ref{cn}--\ref{odeT1} 
can be explicitly calculated as:
\begin{equation}
\hat{c}_n = V_0\frac{j(n)}{n} e^{-bn}\, , \ \ \ 
\hat{t}_m = T(1-e^{-b}) e^{-bm}\, , \ \ \  b > 0.
\label{ctdist}
\end{equation}
Therefore, the expected distribution of t-arrays is exponentially decaying 
with the same rate of exponential decay as the rate of decay of the 
distribution of t-circles (Fig.~\ref{fig:Tarray}). 
On the other hand, the J-factor does not directly influence the size 
distribution of  t-arrays, except through the value of parameter $b$.
A quick comparison of Eqs.~\ref{cdist} and \ref{ctdist} reveals that the 
size distributions of t-circles in the C-model and the CT-model agree although 
the parameter $b$  may be different. Also, both $\hat{c}_n$ and $\hat{t}_m$
are independent of  $\alpha$, i.e., telomere interactions via end invasion 
do not alter the telomere size distributions. The parameter $b$ is uniquely 
determined by the total number of t-repeats (in t-arrays and t-circles) 
in the system 
(Eq.~\ref{totalmass})
\begin{equation}
V_0 \sum_{n=1}^{\infty} j(n)e^{-bn} + T(1-e^{-b}) 
\sum_{n=0}^{\infty} ne^{-bn} = M . 
\label{tma}
\end{equation}
The second sum can be simplified and Eq.~\ref{tma} reduces to
\begin{equation}
V_0 \sum_{n=1}^{\infty} j(n)e^{-bn} + \frac{T}{1-e^{-b}} = M . 
\label{tmb}
\end{equation}
Two  parameters (e.g. $V_0/ M$ and $T/M$) are needed to determine 
the value of $b$ but the value of all three experiment specific parameters 
$V_0, T$ and $M$, is needed for the distribution of $\hat{c}_n$ 
and $\hat{t}_m$ given by Eq.~\ref{ctdist}.  
We refer the reader to Appendix~\ref{sA:CT} for additional information 
on mathematical analysis of coagulation-fragmentation system 
in Eqs.~\ref{cn}--\ref{odeT1} that extends existing results in the literature. 

\subsection{Excess mass}\label{ss:em}
If the total number of t-repeats $M$ in the C-model is too large, 
the equilibrium distribution  in Eq.~\ref{cdist} displays an interesting 
feature---\emph{excess mass}. The key observation is that among 
all equilibria given by Eq.~\ref{cdist} 
there is one of the maximal finite mass $M_{\text{max}}$ (see Eq.~\ref{Mmax}).  
If the initial condition prescribes $M$ too large, $M > M_{\text{max}}$, 
to be accommodated by any finite mass equilibrium  ($b \ge 0$) the extra 
(excess) mass is gradually transfered to higher and higher modes. Thus,
the probability of a presence of  larger and larger t-circles in the sample 
is increasing, while the distribution of telomeres of all sizes converges 
pointwise to the equilibrium of the largest possible mass $M_{\text{max}}$. 
Such a case happens generically  any time the  total count of t-repeats 
exceeds a critical value $M_{\text{max}}$, 
and in mathematical analysis it is connected with  weak convergence. 
For the J-factor given by  Eq.~\ref{jL} that means an algebraic decay 
of the size distribution in Eq.~\ref{cdist}
instead of an exponential decay.
Presence of the numerically observed excess mass  in the C-model was 
also confirmed for the  kinetic rates that lie out of the range studied so far 
by rigorous mathematical analysis of \citet{carr2012}.

\subsection{Species--specific parameters}\label{ss:ssp}
T-repeat length is a 
species-specific parameter that influences telomere looping frequency 
and thus appears to be central for the distribution of the t-circles 
through the J-factor values (see Eq.~\ref{cdist}). On the other hand, 
the diffusive properties of telomeres influence only the overall time scale 
of the size distribution dynamics with no further effect on the equilibrium.

In addition to these well identified dimensional parameters with known 
values the kinetic rates in the
C-model and CT-model depend also on three non-dimensional local 
molecular parameters $\alpha$, $\beta$, and $\gamma$ defined by Eq.~\ref{abg}. 
They characterize relative sizes of the correction factors to the 
diffusion-limited kinetic rates in the individual reactions 
depicted on Fig.~\ref{fig:scheme} that may be species dependent.
The ratio of correction factors to diffusion-limited kinetic rates in 
recombination/end invasion and dissociation of a recombination/invasion 
complex is denoted $\beta$ and $\gamma$, respectively. 
The parameter $\alpha$ depends on the ratio of correction terms to 
diffusion-limited kinetic rates of end invasion and homologous recombination.
The overall time scale of dynamics of the system is determined by  
$g_1\hat{k} = g_1 k /(2 + \beta)$.

\subsection{Experiment--specific parameters}\label{ss:esp}
There are three experiment specific parameters that enter into the model:  
$V_0$, $M$, and $T$. These parameters
 influence the equilibrium distributions $\hat{c}_n$ of t-circles (C-model) 
and of t-circles $\hat{c}_n$ and t-arrays $\hat{t}_n$ (CT-model). 
Note that the total mass measured in the experiment represents only a fraction 
of the t-circles in the sample as the experimental rate of detection 
of telomeres is unknown but certainly smaller than one. Here we assume that the 
detection rate is independent of the size of t-circles, and treat $M$ 
as a free parameter. 

\begin{figure}[htp]
	\centering
	\includegraphics*[width = 0.48\textwidth]{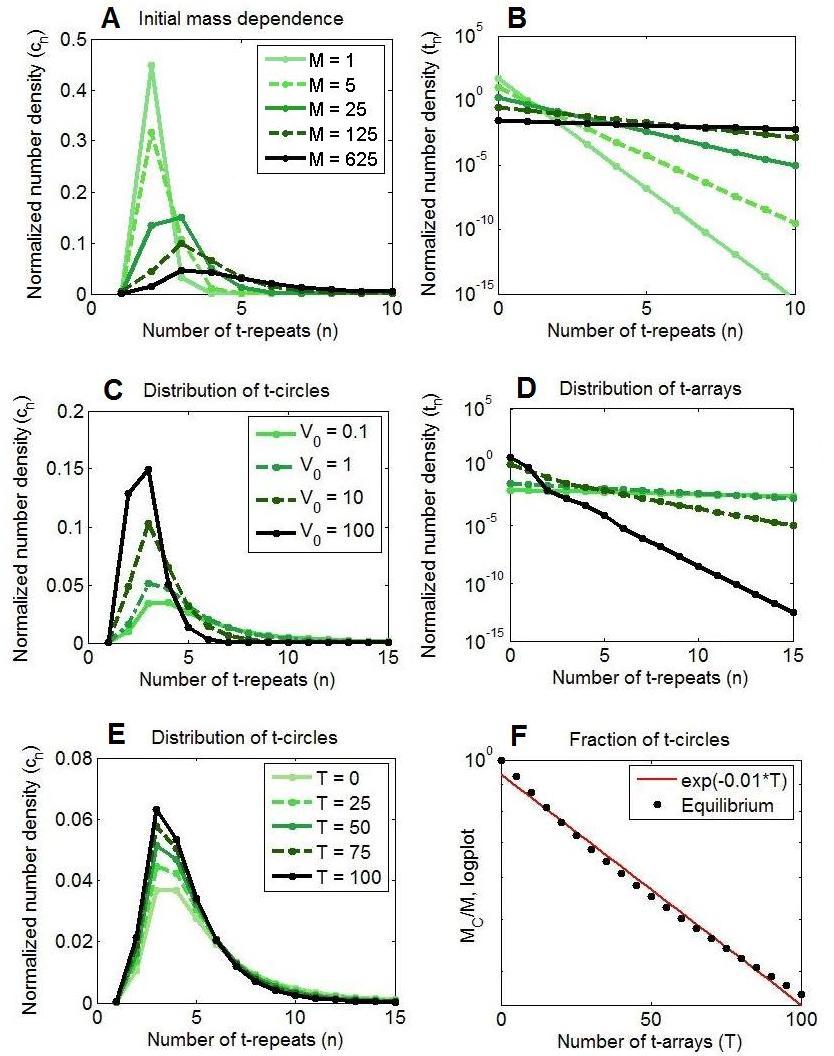}
	\caption{The dependence of the equilibrium t-circle size 
distribution in the CT-model of \emph{C. salmanticensis} 
on the total number of t-repeats $M$ displayed on (A) a decimal 
and on (B) a logarithmic scale ($T = 50$, $V_0 = 1 \mu\text{m}^3$).
The dependence of the equilibrium t-circle size distribution in 
the CT-model of \emph{C. salmanticensis} 
on the sample volume $V_0$ (in units $1 \mu \text{m}^3$) 
displayed on (C) a decimal and on (D) a logarithmic scale 
($T = 100$, $M = 500$). (E) The dependence of the equilibrium t-circle 
size distribution in the CT-model of \emph{C. salmanticensis} 
on the number $T$ of t-arrays ($V_0 = 1 \mu\text{m}^3$, $M= 500$). 
Distributions have a normalized mass for an easier comparison. 
(F) The dependence of the fraction $M_C/M$ of the t-repeats in 
t-circles on the number of t-arrays $T$ in the same model. 
With the increasing $T$ the ratio $M_C/M$ decreases exponentially.}
\label{fig:res3}
\end{figure}

Figure~\ref{fig:res3} AB displays a nonlinear dependence of the equilibrium 
of the t-circle size distribution in the CT-model 
for {\it C. salmanticensis} on the total number $M$ of t-repeats 
in the sample. A larger $M$ implies a shifted mean of the distribution 
towards longer t-circles. Both the distributions of  t-circles 
and t-arrays have exponential tails. 

Similarly, Fig.~\ref{fig:res3} CD shows a dependence of the 
CT-model equilibrium on the volume $V_0$ of the sample. 
Although the equilibrium formula in Eqs.~\ref{ctdist} shows only a linear 
dependence on $V_0$, a different value of $V_0$ also effects the value 
of the parameter $b$ and thus has a nonlinear effect on the t-circle 
size distribution. An increasing sample volume shifts the distribution 
towards smaller circles as the number of the t-repeats in the t-circles 
decreases (with a conserved total number $M$ in the system). 

Moreover, larger $T$ creates an opportunity for more t-repeats to be stored 
in t-arrays than in t-circles thus changing the proportion of the total mass 
in t-circles (see Fig.~\ref{fig:res3} EF).
However, not only the proportion of the total mass in t-circles decreases 
with increasing $T$  but, consequently, 
the shape of the size distribution of t-circles adjusts as well 
(although not significantly). 
According to the numerical observations
\begin{equation}
M_C \approx M e^{-aT}\, , \qquad a > 0\, .
\label{MC}
\end{equation}
We were not able to justify the approximation in Eq.~\ref{MC} rigorously. 

\section{Conclusions and Discussion}
\label{discuss}
A hierarchy of mathematical models of alternative telomere length 
maintenance via recombination on the short time scale was constructed 
based on telomere biophysics. 
A good agreement of the model with experimentally measured size 
distributions of t-circles in yeast 
mitochondria was obtained across various yeast species. The robustness 
of the approach is demonstrated in ability to match the experimental data 
even on the level of the simplest C-model without fitting 
any artificial parameters.
The predicted size distributions of t-circles and t-arrays in both C-model 
and CT-model are characterized by simple algebraic formulae that 
show a high importance of the values of the J-factor for the t-circle 
distributions. 
The size distributions of t-circles are the same 
for both models pointing out the scenario in which the 
dynamics of t-circles is not influenced by t-arrays, i.e., a possible  
universal mechanism of t-circle length maintenance 
independent of t-arrays.  As the model system characterizes the 
recombination dynamics of linear and circular structures in general, 
these properties may be shared by various systems in applications. 

In the rest of  this section we discuss several distinct phenomena 
uncovered  by our analysis that may be tested 
in the biological experiments and lead to an advance in understanding 
of telomere length maintenance. 
We also point out some of the limitations of the proposed model.

\emph{Excess mass.}
Our numerical simulations supported by rigorous mathematical analysis point 
out an interesting phenomenon of excess mass in the C-model responsible for
creation of large t-circles that may be present \emph{in vivo} under the 
condition of sufficiently large density of t-repeats. This mechanism can 
create extremely long t-circles if the cell division halts  but  
cell keeps producing more t-repeats in mitochondria by the rolling circle 
mechanism.  
Whether this phenomenon is biologically relevant and responsible for the 
unusual t-circles observed in experiments 
(e.g., a $C_{10}$ t-circle in \emph{P.~philodendri}, or $C_9$ t-circles in 
\emph{C.~metapsilosis} and \emph{C. parapsilosis}, see Fig.~\ref{fig:fit}) 
is unclear and remains a subject of our further investigations.

\emph{Nonlinear response.}
The rigorous analysis and numerical simulations reveal an interesting feature 
that can be possibly exploited 
in a quantitative estimation of a detection rate of the experimental methods 
(as Southern blot hybridization 
combined with electron microscopy 
\cite{tomaska2000} or quantitative PCR method).  Indeed, the equilibrium 
of the system 
has a nonlinear response to the number of t-repeats in the sample. 
If one interprets the experimentally 
obtained data as relative quantities of measured species of different sizes, 
the model then 
identifies the total number of  t-repeats in the sample by using the location 
and a relative size of the peak 
of the distributions of t-arrays and t-circles. It allows us to relate 
the experimental measurements to an approximate 
number of particular polymers in the sample regardless of the applied technique.

\emph{Long time scale.}
A lack of available dynamical experimental data (time series) motivates 
another key assumption of this study, a very short time scale 
on which the total number of t-repeats in constant. Nevertheless, 
the principal interest lies in the alternative mechanism of 
telomere length maintenance on a longer time scale that includes synthesis 
of t-repeats via rolling-circle mechanism, 
cell division, and subsequently the end-replication problem, and DNA 
degradation. Such a complex model may shed light on many important 
biological question as how many of t-repeats are transfered from a mother cell 
to a daughter cell, what types of telomeres are transfered, how the internal 
cellular clock influences the telomere maintenance, etc., 
but such a study requires additional quantitative data from new experiments. 

\emph{Environmental factors.}
Extensive studies in biophysical literature identify dependences of individual 
factors in kinetic rates 
on environmental parameters as temperature, salinity, or pH. Variations of 
these parameters in 
experiments and the subsequent measurement of the telomere size distributions, 
together with 
the predictions of the model, can eventually both serve as a justification of the 
biophysical components and lead to the further factor-dependent enhancement 
of the model. 

\emph{Single cell vs.~cell population.}
A central assumption of this essay is homogeneity of the environment in 
the experimental sample.  
However, the samples are prepared from multiple cells/mitochondria and 
the resulting distribution corresponds to 
a sum of telomere size distributions in the individual cells/mitochondria. 
A nonlinear dependence of telomere size distributions on the total number 
of t-repeats, 
inhomogeneity of cells in the sample and asynchrony of their phases of cell 
cycles prohibit a more detailed analysis. 
To achieve a better agreement with experimental data
and consequently a better understanding of the telomere length maintenance 
process in a single cell rather than in a cell population, 
single cell measurements need to be obtained experimentally 
or a model with age structured population of cells needs to be developed. 

\emph{J-factor alternatives.}
We have presented our results for the particular choice of the values of 
the J-factor (Eqs.~\ref{jL} and \ref{circJ}). 
However, other choices may be relevant: one may use older Shimada-Yamakawa 
model \cite{Shimada} or to account for recent results  
for short polymers \cite{Han, VaHa}. Moreover, modified circular J-factor 
introduced in \citet{Rippe2001} may be considered as well. By this means 
the experimental measurements of telomere length distributions in yeast 
mtDNA may serve as an alternative to the direct tethered particle methods 
for J-factor estimation. Particularly interesting
would be experimental studies of species that use ALT and have t-repeats 
of length 30–100 bp. Also, simultaneous 
measurement of t-array and t-repeat size distributions would increase the 
robustness of the model, particularly in connection 
with single cell measurements.

\section*{Acknowledgments}
The work was supported by Marie Curie International Reintegration 
Grant 239429 from the European Commission (R.~K.), 
by the VEGA grant 1/0459/13 (R.~K. and K.~B.), and 
by grants APVV-0123-10 and APVV-0035-11 from the Slovak Research 
and Development Agency ({\mL}.~T. and J.~N.).

\appendix

\section{Experimental Setup}\label{sA:setup}
The full description of experimental methods used to obtain t-circle size 
distributions can be found in \cite{tomaska2000} including
the details on yeast strains and DNAs used, DNA isolation, and preparation 
of mitochondria. T-circles isolated from purified mitochondria 
by alkaline lysis were relaxed with DNase I, and aliquots prepared for EM 
were directly adsorbed to thin carbon foils and rotary shadow
cast with tungsten. The samples were subsequently imaged using electron 
microscopy and images were scanned and post-processed 
on a computer to adjust for high contrast. The length of the molecules 
was individually measured (see \cite{griffith1999mammalian} 
for more details). 

\section{Quasi-Steady State Approximation and Reduction to the CT-model} 
\label{sA:QSSA}
Time scale separation allows to reduce a reaction kinetics system by the 
means of the quasi-steady state approximation (QSSA)
\citep{Atkins,Eigen}
However, the QSSA can often be misleading and the conditions for its validity 
need to be checked carefully. We refer the reader to works of  
\citet{schnell2000} 
and \citet{flach2006} for examples of wrong use of the QSSA. 

\citet{segel1988} and \citet{SegelSlemrod} were the first to point out the 
connection of the QSSA in the Michaelis-Menten enzyme kinetics to the mathematical
 problem of singular perturbation.
They also derived the proper conditions necessary for the validity of the QSSA 
for simple enzyme
kinetics. These techniques for irreversible enzyme kinetics were extended 
beyond the traditional
QSSA by \citet{SM2002, SM2003} to cover the regimes where the QSSA fails 
(see also \citep{tzafriri2007}).
Further (formal) extensions of the QSSA and related techniques for the 
reversible enzyme kinetics
and bimolecular reactions were derived by \citet{tzafriri2004,TE2004}.

The existing theory allows to formulate conditions for the validity of the
QSSA for the individual reactions (R4), (R5), and (R6) 
(see Fig.~\ref{fig_reactions}). For the three step reversible reactions (R1) and (R2) and 
the partially reversible three-step reaction (R3) it is possible to formally 
derive the conditions by using an approach analogous to \citep{TE2004}. 
However, the theory is not applicable for the infinite system of reactions 
(R1)--(R6) with telomeric structures of arbitrary size because the time scales
in the system are intertwined.  Any rigorous result justifying the QSSA 
for the infinite system
(R1)--(R6) would be a significant contribution to the theory of the QSSA 
as even for the system (R1)--(R6) 
of finite size (with limited maximal length of the telomeric structures) 
the right conditions
for validity of the QSSA have not been yet established.

Nevertheless, our numerical simulations indicate the validity of the QSSA 
in the system (R1)--(R6) that allows to eliminate all 
the intermediate products 
of all sizes in the system schematically described on  Fig.~\ref{fig_reactions}.
Therefore we use the QSSA for the whole system (R1)--(R6) 
and replace the ordinary differential equations describing
the temporal dynamics of populations of recombination 
complexes and end invasion complexes
by the equilibrium conditions imposing algebraic laws 
for the populations of complexes.
The simplified system referred to as the  
CT-model consist of the mass action kinetic reactions in Eqs.~\ref{eq_CC}--\ref{eq_TT}
with the kinetic rates given by 
\begin{eqnarray*}
k^{CC} (C_m,C_n)& = &  k_1\cdot \frac{k_3k_5}{k_2k_4+k_2k_5+k_3k_5} \, , \\
k^{C} (C_m,C_n)& = & k_6\cdot \frac{k_2k_4}{k_2k_4+k_2k_5+k_3k_5} \, ,\\
k^{TC} (T_m, C_n)& = & k_7\cdot \frac{k_9k_{11}}{k_8k_{10}+k_8k_{11}
+k_9k_{11}}\\
& & +k_{21}\cdot \frac{k_{23}}{k_{22}+k_{23}}\, ,\\ 
k^{T} (T_m,C_n)& = & k_{12}\cdot \frac{k_8k_{10}}{k_8k_{10}+k_8k_{11}
+k_9k_{11}} \\ & & 
+ k_{18}\cdot \frac{k_{20}}{k_{19}+k_{20}}\,,\\ 
k^{TT} (T_m, T_n, T_p)& =&  k_{13}\cdot \frac{k_{15}k_{17}}{k_{14}k_{16}
+k_{14}k_{17}+k_{15}k_{17}} \\ & & +k_{24}\cdot \frac{k_{26}}{k_{25}+k_{26}}\, .
\end{eqnarray*}
A straightforward calculation yields the simplified kinetic rates
\begin{eqnarray*}
k^{CC} (C_m,C_n)& = &  mn \hat{k} g_1 \, , \\
k^{C} (C_m,C_n)& = & (m+n)  j(m,n) \hat{k}g_1  \, ,\\
k^{TC} (T_m, C_n)& = &  n  (m + \alpha)\hat{k} g_1 \,,\\ 
k^{T} (T_m,C_n)& = & \left(m j(n) + \alpha j^{\ast}(n)\right) \hat{k} g_1\,,\\ 
k^{TT} (T_m, T_n, T_p)& =&   
\left( \min(m,n,k,m+n-p) + \alpha\right) \hat{k} g_1
\end{eqnarray*}
that reduce after rescaling of time and an assumption $j(n) = j^{\ast}(n)$ to Eq.~\ref{kTCr}.

\section{Mathematical formulation of the CT-model and the C-model}
\label{sA:CT}
The system of reaction kinetics in chemical reactions in Eqs.~\ref{eq_CC}--\ref{eq_TT} 
can be mathematically formulated as two 
coupled sets of infinite number of ordinary differential equations in Eqs.~\ref{cn}--\ref{odeT1}. 
The system describes time evolution of the concentrations $c_m$ and 
$t_n$ of the populations $C_n$, $n > 0$,  and $T_n$, $n\geq 0$, respectively, in the sample solvent.
The Gaussian circular J-factor satisfying the detailed balance condition in Eq.~\ref{circJ}
allows for an explicit equilibrium solution of the system of 
Eq.~\ref{cn}--\ref{odeT1} 
given by Eq.~\eq{ctdist1}
The free positive parameter $b$  is uniquely 
determined by the total initial number of t-repeats $M$ in the sample.
Note that while the t-circle size distribution is determined (up to 
a factor $n$) by the product of the J-factor with the decaying exponential, 
the t-arrays distribution is exponentially decaying with 
the same exponential rate.  

The CT-model does not fit into any traditionally considered classes of systems 
of differential equations and we are not aware of any known general 
analytical results for the system of Eqs.~\ref{cn}--\ref{odeT1}. 
On the other hand, the reduced C-model formulated as an infinite system of Eqs.~\ref{odeC2}
that describes the evolution of the population $c_m$ of t-circles of size 
$m$ (full t-repeats) falls into a class of typically studied systems of 
ordinary differential 
equations --- \emph{the coagulation fragmentation systems}.
The parameters $k_{m,n}^{C}$ and $k_{m,n}^{CC}$ are called 
the fragmentation and the coagulation kernel, respectively. 

The theory of coagulation-fragmentation systems concerns rigorous 
mathematical studies of either discrete systems of the form \eq{odeC2} 
or their continuous counterparts that allow non-integer sized particles in which 
the vector $(c_1,c_2, \dots)$ 
 is replaced by a nonnegative real function 
$c = c(m)$, $m\ge 0$, and the sums on the right hand of Eq.~\ref{odeC2} are replaced 
by integrals. In general such systems 
describe the dynamics of cluster growth with applications in biology, 
material science, or atmospheric physics (see \citep{CFthesis} 
for an extensive survey). 

Since there is a large amount of literature on the subject
we only  provide here a short list of publications and recommend an interested 
reader to search for further references within. The mathematical 
treatment of the problem in Eq.~\ref{odeC2} in the context of 
physical chemistry can be traced back to an influential paper of 
\citep{AizBak} and its predecessor \citep{BakBak} where a survey 
of older literature can be found. \citet{AizBak} studied constant 
interaction kernels and proved uniqueness of solutions of Eq.~\ref{odeC2} 
and their convergence to an equilibrium. Later on, \citet{BallCarr} 
studied various types of the discrete coagulation-fragmentation models 
for a wider class of additive kernels. \citet{daCosta} 
proved existence and uniqueness of density conserving solutions for 
models with strong fragmentation in the sense of \citet{Carr}. More 
recently, \citet{Fournier2004} generalized the asymptotic results to 
the case of strong fragmentation. Their analysis requires a relative 
smallness of the initial data $M= M_C$.

In \citep{Carr2012} the system in Eq.~\ref{odeC2} is considered 
with three different choices for coagulation and fragmentation kernels. 
First, in the case of constant J-factor, the kernels are given by 
\begin{equation}
k_{m,i}^{C} = m+i, \qquad
k_{m,i}^{CC} = mi.
\label{caseIkernels}
\end{equation}
In that case the kernel only accounts for the combinatorial part 
of the real biophysical t-circle interactions.  
In this simple case the system is exactly solvable by the means 
of Laplace transform and it is easy to establish that 
an arbitrary initial data converge (uniformly on finite sets) to the 
unique equilibrium given by  
\begin{equation}
c_m = \frac{e^{-bm}}{m}\, , \qquad \quad b > 0,
\label{caseIeq}
\end{equation}
where the free positive parameter $b$  is uniquely determined 
by the initial data.  Moreover, the distribution converges to the 
equilibrium exponentially in time. 

Due to the fact that the circular J-factor satisfies the detailed 
balance condition in Eq.~\ref{circJ}  the
equilibria of the system can be explicitly expressed as
\begin{equation}
c_m = \frac{j(m)}{m}e^{-bm}\, , \qquad \quad b \ge 0,
\label{caseIIeqgen}
\end{equation}
providing a very good agreement with the experimental data. 
Carr mathematically analyzed the particular case of the J-factor 
of the Gaussian polymer model with $j(m) = m^{-3/2}$ 
(after a normalization, see Eq.~\ref{jG}). Then  
\begin{equation}
k_{m,n}^{C} = \frac{(m+n)^{5/2}}{(mn)^{3/2}}, \qquad
k_{m,n}^{CC} = mn, 
\label{caseIIkernels}
\end{equation}
In that case the set of equilibria is given by 
\begin{equation}
c_m =  \frac{e^{-bm}}{m^{5/2}}\, , \quad \qquad b \ge 0,
\label{caseIIeq}
\end{equation}
and there is an equilibrium of the maximal (critical) finite mass  
$$
M_{\text{max}} = \sum_{m=1}^{\infty} m^{-3/2}
$$ 
(we set $V_0 = 1$ here for simplicity). 
Then all the solutions with the initial mass smaller than  $M_{\text{max}}$ 
converge (uniformly on finite sets) to the appropriate equilibrium while 
the solutions with the initial mass larger or equal than $M_{\text{max}}$  
converge (pointwise) to the equilibrium in Eq. \ref{caseIIeq} with $b=0$, 
and the excess mass is moving to higher and higher modes indicating 
weak convergence of the solution to the equilibrium of the critical 
mass in an appropriate functional space. Note that the scaling of the 
critical size distributions in Eq.~\ref{caseIIeq} (for $b=0$) 
is the same as in \citep[Eq.~7]{bennaim} illustrating the generic 
nature of the phenomenon. 
\citet{Carr2012} also considered more general case that accounts 
for size-dependent diffusion of the telomeres but the problem due 
to the change of the structure of the reaction rates posed 
a difficult unsolved mathematical problem. 

An interesting feature of the CT-model (Eqs.~\ref{cn}--\ref{odeT1}) 
under the assumption of the circular J-factor with
the detailed balance condition is that the algebraic form of 
its equilibrium given by Eq.~\ref{ctdist} is in agreement 
with the equilibrium (Eq.~\ref{caseIIeqgen}) of the reduced C-model. 
The general reason for this fact is the compatibility of the
kinetic reactions described by Eq.~\ref{eq_CC}--\ref{eq_TT} 
connected to preservation of the mass (the number of t-repeats) 
in each individual reaction. 
However, if the circular J-factor is modeled using the theory of \citet{Rippe2001}, 
the equilibrium conditions 
forced by individual reaction in Eq.~\ref{eq_CC}--\ref{eq_TT} 
are not compatible and there is no such simple formula for the equilibrium of the system.

\begin{widetext}
\centering
\begin{table}[htp]
\begin{tabular}{|p{8.5cm}|p{9cm}|}
\hline
{\bf Biophysical parameters} & Value \\ \hline
Typical length of a basepair &  $\xi = 0.34$ nm (width 2 nm) \citep{moran} 
\\ \hline
Typical yeast cell size (\emph{S. cerevisiae}) & $10 \text{$\mu$m}$ (diameter), 
20--160 $\text{$\mu$m}^3$ (volume) \citep{moran} 
\\ \hline
Estimated number of mtDNA copies
in a haploid yeast cell ({\it S. cerevisiae}) & 20-50 (10-20\% of total DNA) 
\citep{Williamson} \\ \hline
Cell cycle time budding yeast & 70--140 min \citep{moran} \\ \hline
Mitochondrial genome size, telomeric tandem repeat length \emph{C. salmanticensis} 
&  $25\,718 + (m+n)\times 104$ bp 
\citep{nosek1995linear,tomaska2000,valach2011} \\ \hline
Mitochondrial genome size, telomeric tandem repeat length \emph{P. philodendri}
 & $26\,487 + (m+n) \times 288$ bp \citep{nosek1995linear} \\ \hline
Mitochondrial genome size, telomeric tandem repeat length \emph{C. metapsilosis}
 & $23\,062 + (m+n) \times 620$ bp \citep{kosa2006} \\ \hline
Mitochondrial genome size, telomeric tandem repeat length \emph{C. parapsilosis}
 &  $30\,923 + (m+n) \times 738$ bp \citep{nosek1995linear} \\ \hline
Length of the single-stranded 5$'$ overhang & $110 \times \xi$ nm (110 nt) 
\citep{nosek1995linear} \\ \hline
Estimated typical length of a hotspot & $l_{\text{HS}}= 10$ bp \\ \hline
Bending and torsional persistence length of DNA as a homogeneous polymer &  
$\el_p = 50$ nm \citep{Wilson}\\ \hline
Boltzmann constant & $k_B = 1.3806503 \times 10^{-23} \text{m}^2 \text{kg} 
/ \text{s}^2 \text{K}$ \\ \hline 
Absolute temperature & $T_{\text{a}} = 273.15 K$ \\ \hline
Solvent viscosity & $\hat{\nu} =1.2 \times 10^{-9} \text{kg} / \text{$\mu$m}\, 
\text{s}$ \citep{robertson2006}\\ \hline
\end{tabular}
\caption{Biophysical parameters}
\label{tab:BP}
\end{table}
\end{widetext}



\end{document}